%% file: main.tex
\documentclass{article}
\usepackage{natbib}
\usepackage{authblk}

\usepackage{fix-cm}
\usepackage{amssymb}
\usepackage{amsmath}
\usepackage{graphicx}
\usepackage{subfigure}
\usepackage{multicol}
\usepackage[normalem]{ulem}
\usepackage{bm}
\newcommand{\DC}[1]{}%{\textcolor{red}{[Dan note: #1]}}  %used for comments 

\usepackage{amsthm}

\usepackage{color}
\usepackage{xcolor}
\usepackage{algorithm}

\usepackage{hyperref}
\hypersetup{
    colorlinks=true,
    linkcolor=blue,
    filecolor=magenta,      
    urlcolor=cyan,
    citecolor=blue
    }

\newcommand{\as}[1]{} %{{\color{violet}[{\tiny A:~}#1]}}
\newcommand{\EE}{\text{E}}

\newcommand{\PP}{\text{P}}
\newcommand{\rset}{\mathbb{R}}

\newcommand{\sphere}{\mathbb{S}}
\newcommand{\ball}{\mathbb{B}}
\newcommand{\Pareto}{\mathrm{Pareto}}
\newcommand{\Span}{\mathrm{Span}}

\newcommand{\proj}{\Pi}

\newcommand{\vertiii}[1]{{\left\vert\kern-0.25ex\left\vert\kern-0.25ex\left\vert #1
    \right\vert\kern-0.25ex\right\vert\kern-0.25ex\right\vert}}

\newcommand\wto{\xrightarrow[]{\text{w}}}

\newtheorem{proposition}{Proposition}
\newcommand{\argmin}{\operatorname{argmin}}

\newcommand{\tr}{tr}

\newcommand{\TPDM}{{\Sigma}} %% non-centered covariance matrix defined in the introduction
\newcommand{\Thet}{\Theta} % X/\|X\|
\newcommand{\T}{^{\top}}
\newcommand{\thetab}{\boldsymbol{\theta}}
\newcommand{\pto}{\longrightarrow}

\begin{document}
\title{ Principal Component Analysis for Multivariate Extremes
}%\label{ch:pca}
\author{Daniel Cooley$^{(1)}$, Anne Sabourin$^{(2)}$ \& Troy Wixson$^{(3)}$ \\~\\
  \textit{Chapter 11 in:} ``Handbook of Statistic of Extremes'', \\
  \emph{ Editors: Miguel de Carvalho, Raphaël Huser, Philippe Naveau, Brian Reich.} \\~\\~\\
{\normalsize$(1)$: Department of Statistics, Colorado State University, Fort Collins, Colorado\\
  $(2)$: Université Paris Cité, Université Paris Saclay, ENS Paris Saclay, CNRS, SSA, INSERM, Centre Borelli, F-75006, Paris, France \\
  $(3)$: Department of Mathematics and Statistics, University of Massachussetts, Amherst.}}
\maketitle
%\vspace{-0.6cm}
%\begin{shadebox}\vspace{-0.4cm}
  \begin{quote}
      \textbf{Outline:} \it While previous chapters focused on modeling multivariate extremes, we now turn to dimension reduction methods. This chapter explores ways to reduce the dimensionality of the data while preserving key information relevant to the analysis of multivariate extreme values.
\end{quote}\vspace{-0.5cm}
%\end{shadebox}~\vspace{-1.6cm}
~
\section{Introduction}\label{sec:intro} 

% \textcolor{blue}{From Anne: Historical bottlenecks (Heavy-tails $\to$
%   no guarantee of high order moments $\to$ lack of statistical
%   guarantee for PCA applied to raw (untransformed) data.  Historical
%   works.}
%\input{diffintro}
\input{introduction}
%\as{put the algorithm here}

\section{PCA for Extremes}\label{sec:backgroundPCARV} ~\vspace{-0.7cm}
\input{backgroundPCA-RV}
\section{Extremal PCA: Setup and Estimation}\label{sec:probAndEst}~\vspace{-0.7cm}\index{Dimension reduction in multivariate extremes!extremal PCA (Principal Component Analysis)} 
% DS:  \\
% (1) define projection at bottom of page 911.  Question:  $\theta$ lives on unit ball.   Does $\Pi_V\theta$ live on the unit ball?  Does it need to? \as{ no it does not, and it does not need to}\\
% (2) page 912:  Define risk $R_{\infty}(V)$ squared error of limiting measure between (?? any angular component $\theta$ and its projected value $\Pi_V\theta$  ??)
% \as{ Indeed it is not so clear in the paper: let's try again:
%   $$
% R_\infty(V) = \EE[ \|\Pi_V(\Theta_\infty) - \Theta_\infty\|^2 \|], 
% $$
% where $\Theta_\infty = \theta(X_\infty)$ and $X_\infty$ follows the limit distribution of extremes $P_\infty(\,\cdot\,)=  \lim_t \PP( X/t \in \,\cdot\, |\|X\|>t)$. I suggest we write it like this, not like in the paper, if it looks clearer to you 
% }\\
% (3)  page 913, Lemma 2.1:  Say that generally $\Sigma = E(YY^\top)$ minimizes projection error. \as{yes, this is the general idea}
% Define covariance operator of extreme angles.  \as{already in the introduction (? if you agree)}

\subsection{Probabilistic Setup}\label{sec:probaSetting}
\input{probaPCAExtremes}

% \as{ previous notes below:}\\
% definitions - relationships with asymptotic independence and a sparse support of the exponent measure/anguar measure - Prove (This is simply weak convergence of $P_t$ to $P_\infty$ together with teh fact that $\theta(x)$ is a boounded, continuous mapping) the  identity
% $$
% \TPDM (~\eqref{eq:TPDM}) = \EE[\Theta_\infty\Theta_\infty^\top]
% $$
% (the covariance matrix of the limit is the limit of the covariance matrices)

\subsection{Estimation}\label{sec:estimation}
\input{statsDS}
\section{Principal Components and Interpretation}
\label{sec:extPCs}

\input{cooleyThibaud}
\section{Applications} \label{sec:applis} ~\vspace{-0.7cm}
\input{applicationFinance} \label{sec:appFinance}

\input{applicationRainfall} \label{sec:appRainfall}

%note:  begin with explanation of the geometry of rainfall?

\section{Perspectives}\label{sec:perspectives}

\input{perspectives}

% \begin{table}
% \tabletitle{Now we are engaged $(a_g^a)$ $\big(a_g^a\big)$ in a great civil war, testing whether that nation, or any nation so conceived.}\label{t:civil_war}
% \begin{tabular}{lccc}
% \tch{Scene}    &\tch{Reg. fts.} &\tch{Hor. fts.} &\tch{Ver. fts.}\\
% Ball &19, 221 &4, 598   &3, 200\\
% Pepsi$^a$&46, 281 &6, 898 &5, 400\\
% Keybrd$^b$   &27, 290 &2, 968 &3, 405\\
% Pepsi    &14, 796 &9, 188 &3, 209\\
% \end{tabular}
% \end{table}

% \begin{figure}[htb]\label{f:cat}
% \includegraphics[width=200pt, height=200pt]{Ch11/LaTeX/figures/cat.pdf}
% \caption[Short figure caption]{Figure caption goes here. Figure caption goes here. Figure caption goes here. Figure caption goes here. Figure caption goes here. Figure caption goes here.}
% \end{figure}

% \begin{figure}
% \begin{center}
% \subfigure[\label{f:f8a}]{\includegraphics[width=0.4\textwidth]{Ch11/LaTeX/figures/cat.pdf}}
% \subfigure[\label{f:f8b}]{\includegraphics[width=0.4\textwidth]{Ch11/LaTeX/figures/cat.pdf}}
% \end{center}
% \caption[Here are two cats]{The two cats are named (a) Josie and (b) Tiger.}
% \end{figure}

%\input{classicalPCA}

% \mainmatter

\bibliographystyle{plain}%nat}%apalike}
\bibliography{bibtexPCA}

% \printindex
% \cleardoublepage

 \end{document}

%% file: introduction.tex
%\as{I've re-organised a little bit the argument, I now  start with pointing out differences and end with similarities}

% \DC{The editor and reviewer 2 seem to suggest beginning with gentle and applied introduction to PCA.  We could begin here with something along the lines of "Informally, (non-extreme) PCA finds the directions of greatest imporatance, as measured by variance.  Then, list of tasks:  dimension reduction, interpretation, widely used by practitioners and largely exploratory...  Could add the simple 2-d visual that the editor suggests.  Conclude by setting up for difference from extremes:  Traditional PCA looks from the center outward, limited in ability to capture heavy tails (covariance matrix requires 2 moments, but statistical guarantees in fact require 4), best suited for data that are roughly elliptical.  Could potentially pull some material from file "classicalPCA.tex" which Dan wrote but which was discarded in favor of the presentation that appeared in \S 1.2.1 of original submission.  Rebrand 1.2.1 as a more formal/statistical(?) development of PCA.  Reviewer 2 suggests moving discussion of angular measure forward.  Also suggests mentioning centering, screeplot, maybe covariance vs correlation matrix }

% \DC{For now, just throwing out some text.  It all needs to be thought about more.}

%pca fluff
\paragraph{Background on Principal Component Analysis}\index{Dimension reduction!PCA (Principal Component Analysis)}
Principal component analysis (PCA) is a method widely used by practitioners for learning features of high-dimensional data \cite{greenacre2022principal}. It is a dimension reduction technique that represents the data in lower dimensions, often with the aim of exploratory analysis or visualization. PCA can also be used as a data preprocessing step, for instance in regression analysis.
While PCA is familiar and commonplace for understanding behavior in the data's `bulk', only recently have similar methods been proposed for understanding high-dimensional extremes.
The aim of this chapter is to review and compare recent approaches for extremal PCA.

%gentle introduction random variable POV
To provide context, we begin with a brief and somewhat informal presentation of classical PCA.
A more formal presentation based on projection will be given in \S \ref{sec:PCA}, as this will be necessary for the development of extremal PCA.
Classical PCA can be understood in terms of maximizing the variance of uncorrelated linear combinations of a $D$-dimensional random vector $\mathbf{X}$ with finite second moments \cite[\S 8.2]{johnsonwichern2007}.

Let $\mathbf{a}_j \in \mathbb{R}^D$ be unit-norm vectors, i.e., $\| \mathbf{a}_j \| = 1$, for $j = 1, \dots, D$. The \index{Dimension reduction!principal component}\textit{principal components} of $\mathbf{X}$ are defined as $$Z_j =  \mathbf{a}_j\T  \mathbf{X}, \quad j = 1, \dots, D,$$ where $ \mathbf{a}_1$ is such that $\text{Var}(Z_1)$ is maximized, and for $j = 2, \ldots, D$, $\mathbf{a}_j$ is such that $\text{Var}(Z_j)$ is maximized subject to $\text{Cov}(Z_j, Z_k) = 0$ for $k = 1, \ldots, j-1$. 
It is well known that the $\mathbf{a}_j$'s follow by the eigendecomposition of the covariance matrix $\Sigma \equiv \text{Cov}(\mathbf{X})$. Indeed, the eigendecomposition of $\Sigma$ yields eigenpairs $\{(\mathbf{a}_j, \lambda_j)\}_{j = 1}^D$, where the eigenvalues are ordered as $\lambda_1 \geq \cdots \geq \lambda_D \geq 0$. Furthermore, $\text{Var}(Z_j) = \lambda_j$, and so $\lambda_j / \sum_{j = 1}^D \lambda_j$ is the percentage of the total variance of $\mathbf{X}$ explained by the principal component $Z_j$.
%If only a subset $Z_1, \ldots, Z_p, p < d$ of the principal components are retained, then these represent $ \sum_{j = 1}^p \lambda_j/  \sum_{j = 1}^D \lambda_j$ of the total variance found in $ X$.

\begin{figure}[htb]
\includegraphics[scale = .6]{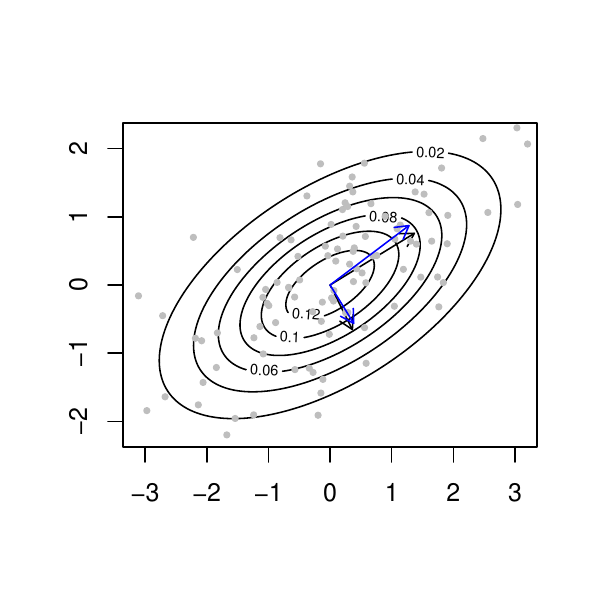}
\includegraphics[scale = .6]{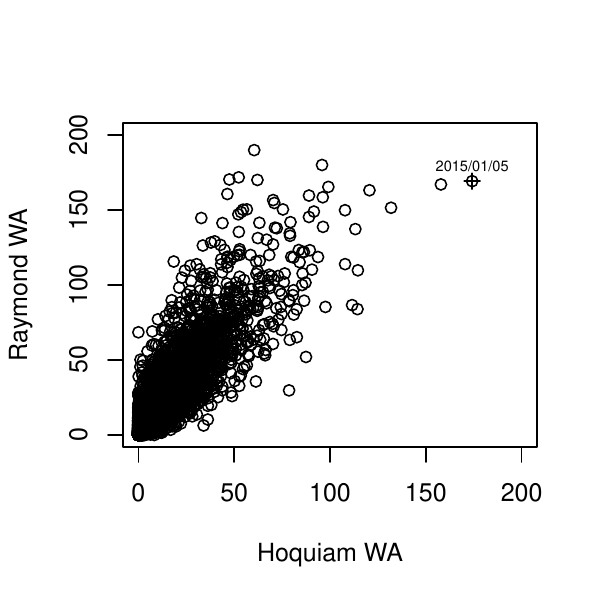}
\caption[Short figure caption]{Left:  Contour of the density of a bivariate Gaussian distribution; black arrows give the direction and standard deviation of the principal components $Z_1$ and $Z_2$.  Shown also are 100 realizations from the distribution and the blue arrows give the direction and standard deviation of the estimated principal components.  Right: daily precipitation measurements (mm) at two locations in Washington State, USA;  the labeled point corresponds to a day of a known flooding event.}
\label{fig:classicalPCA}
\end{figure}

The left panel of Fig.~\ref{fig:classicalPCA} illustrates PCA for a two-dimensional Gaussian distribution with covariance matrix 
$$\Sigma = 
  \left( 
    \begin{array}{c c}
    2 & .8\\
    .8 & 1
    \end{array}
  \right).
$$ 
The black arrows correspond to  $\sqrt{\lambda_1} \mathbf{a}_1$ and $\sqrt{\lambda_2} \mathbf{a}_2$ giving the direction and standard deviation of the principal components $Z_1$ and $Z_2$.

%jump to data
In practice, the covariance matrix is not known and must be estimated.
Starting with independent and identically distributed realizations,  $\{\mathbf{x}_i\}_{i = 1}^n$, the \index{Dimension reduction!sample principal component}\textit{sample principal components} are obtained by taking an eigendecomposition of the sample covariance matrix $\hat \Sigma$.
Let $\hat \lambda_1 \geq \cdots \geq \hat \lambda_D$ be the eigenvalues and 
$\hat{\mathbf{a}}_1, \ldots, \hat{\mathbf{a}}_D$ be the corresponding eigenvectors 
%$\hat U_k = (\hat{ u}_1, \ldots, \hat{ u}_k)$ be the matrix of the first $k \leq d$ ordered eigenvectors 
obtained from $\hat \Sigma$.
The sample eigenvalues and eigenvectors are typically interpreted as explained above, but development of the statistical properties of these estimates is  involved and requires conditions such as the existence of fourth moments
(\cite{anderson1963asymptotic,greenacre2022principal,joliffe}).
The blue arrows in the left panel of Fig.~\ref{fig:classicalPCA} correspond to $\sqrt{\hat \lambda_1} \hat{\mathbf{a}}_1$ and $\sqrt{\hat \lambda_2} \hat {\mathbf{a}}_2$, where $\hat \Sigma$ was estimated from 100 realizations from the aforementioned Gaussian distribution.

Since the eigenvalues are ordered, the dimension reduction aspect of PCA arises from the practitioner selecting a number of $d < D$ principal components to retain.
%Letting $\hat U_p = (\hat{ u}_1, \ldots, \hat{ u}_p)$ be the matrix of the first $p$ ordered eigenvectors and $\hat { \mu}$ be the estimated mean vector, the vector of sample principal components $ z_i = \hat U_p^{\T} ( x_i - \hat { \mu}_i)$ can be obtained for each observation $i = 1, \ldots, n$.
Let $\hat{\mathbf{a}}_1, \ldots, \hat{\mathbf{a}}_d$ be the first $d$ eigenvectors and $\bar{\mathbf{x}}$ be the estimated mean vector, and define the $j$th sample principal component of the $i$th observation as $z_{i,j} = \hat{\mathbf{a}}_j^{\T} (\mathbf{x}_i - \bar{\mathbf{x}})$, for $j = 1, \ldots, d$.
The collection of these first $d$ principal components represent $ \sum_{j = 1}^d \hat{\lambda}_j/  \sum_{j = 1}^D \hat{\lambda}_j$ of the total variance found in the data.
The lower-dimensional sample principal components can be investigated more easily than the high-dimensional original data; for example, plots of the leading sample principal components can identify outliers or other interesting observations.
Any observation $\mathbf{x}_i$ can be partially reconstructed from its principal components:  
%$\tilde x_i = \hat U_p  z_i + \hat { \mu}_i$ 
$\tilde{\mathbf{x}}_i = \sum_{j = 1}^d z_{i,j} \hat {\mathbf{v}}_j$ can be considered as the most effective $d$-dimensional representation of $ \mathbf{x}_i$, and if $d = D$, then $ \mathbf{x}_i$ is reconstructed exactly. 
There exist various rules-of-thumb and hypothesis tests for selecting $d$.
However, $d$ is often chosen subjectively, the choice is informed by a scree plot\index{Dimension reduction!scree plot} (plot of eigenvalues in decreasing order) or other diagnostic measure, and the choice of $d$ often depends on the task at hand.
\paragraph{Principal Component Analysis for Multivariate Extremes---Why and How}
As it does not focus on the data's extremes, classical PCA is a poorly suited tool for understanding extremal behavior.
Rather than focusing on the tail, the covariance matrix describes dependence from the center of the distribution outward.
Furthermore, although PCA does not require an assumption of Gaussianity, it is most useful when the data are roughly elliptical; i.e., the density's contours are roughly elliptical. 
Data being analyzed for extreme behavior are often strongly non-Gaussian.
%The most disruptive extreme events often have distributions with heavy tails, such as in finance and many natural phenomena.F
The right panel of Fig.~\ref{fig:classicalPCA} shows rainfall data at two nearby locations in the state of Washington, USA.
Clearly these nonnegative data are not elliptical.
Furthermore, like many phenomena where extreme behavior can be highly disruptive, this data will be found to have heavy tails in \S \ref{sec:precip}.
Finally, classical PCA does not focus attention in the direction associated with risk; here, our interest lies in understanding the highest precipitation values (found in the upper right of the figure) which likely cause flooding.
Covariance looks from the center outward and characterizes the strength of dependence (in both directions) with a single number.
%\DC{Using this last idea to help set up for the editor comments about direction.}

%\DC{This 'gentle' introduction has focused on the covariance matrix rather than the uncentered matrix of second moments which is more closely related to the extremal PCA.  I think this makes sense here, as I think most practitioners are going to be familiar with the covariance matrix and most applied PCA begins with a covariance matrix rather than an uncentered matrix.  It also provides an opportunity to point out this difference when we turn to extremal PCA.  But it will need to be explained a bit more.}

%#######

%\as{general intro/pitch:}\\
Developing PCA analogues specific for analyzing high-dimensional extremes is challenging for a couple of reasons.
First, a basic tenet of extreme value analysis is to only analyze a small subset of extreme data in order to focus solely on the extreme behavior.
The scarcity of extreme data combined with the curse of dimensionality suggests high dimensional extremes methods are hard.
%A reasonable concern is that the scarcity of data inherent to EVA would prevent any success in attacking high dimensional problems. 
Second, extremes methods need to be applicable to heavy tailed data, and the requirement of existence of fourth moments so to obtain statistical guarantees for PCA is a prevalent obstacle.
Theoretical guarantees typically concern
  the   reconstruction error (\cite{blanchard2007statistical,koltchinskii2000random,koltchinskii2016new,shawe2005eigenspectrum,reisswahl2020}) or the approximation error for the eigenspaces of the covariance matrix (\cite{zwald2006convergence}).

Recent works have shown that the  technical difficulties associated with heavy tails may be
overcome, and that a sensible notion of PCA may indeed be designed for modeling multivariate extremes, with statistical guarantees and interpretations  paralleling the classical
ones. 
This chapter synthesizes two recent
approaches of PCA for extremes that have been developed independently,
namely the works of \cite{CT} and \cite{DS}. %Although 
For the sake of simplicity, we limit ourselves in this chapter to the $D$-dimensional context although an extension  to functional PCA in the setting of functional  data analysis for observations valued in Hilbert spaces has recently been developed in \cite{clemenccon2023concentration}.

%\as{todo mention also \cite{clemenccon2023regular}, maybe later on?}
%on the one hand   with a  follow-up application \cite{rohrbeck2023}, and that of \cite{DS} on
%the other hand. %Despite the common ground between~\cite{CT} and~\cite{DS},

%\DC{Editor (items 3 and 4) seems to suggest that we more directly address directionality of extremes here.}
\paragraph{On Two Approaches Rooted in Regular Variation and PACE \cite{CT,DS}}
%\vspace{-0.5cm} \strut \hfill \textcolor{gray}{\footnotesize \cite{CT, DS}}\\
%\noindent
At first view, the respective foci and results of~\cite{CT} and~\cite{DS} appear quite different. %,   main focus and the nature of
%their results are different:
\cite{CT} is largely focused toward applications and transposing the `standard' interpretations of PCA to the context of extremes;  
another contribution of  \cite{CT} is to propose a framework aimed at handling positive observations while keeping the technical advantages of a vector space, which is achieved through `transformed-linear' arithmetic and the $\tau$-transform.
On the other hand, \cite{DS} disregards any positivity constraint  but establish statistical guarantees regarding the PCA algorithm they consider, in the form of non-asymptotic upper bounds on the reconstruction error. 
In terms of interpretation, their purpose is rather to establish a connection between the main eigenspaces of the covariance matrix of extreme angles, and sparsity patterns among extremes. In other words their focus is on identifying `good' dimension reduction spaces in terms of reconstruction error, rather than on interpreting the eigenvectors of the covariance matrix.  

However both approaches  share in common similar
fundamental probabilistic objects, in particular regular variation assumptions (see \S \ref{sec:regvar}) and  second moment matrices involving angular components of extreme observations.
%  In this chapter  we consider applying 
%  standard
Also, in practice both  approaches consist of applying PCA  to a re-scaled  version of radially thresholded observations, thereby focusing on extremes. 
The aim of  this chapter is to present a review and interlink contribution of both works, which we believe complement each other.

Ultimately, we aim to use the results of \cite{DS} to justify an extremal PCA algorithm, which we name {PACE} (PCA of the Angular Components of Extremes), and then to link it to the work of \cite{CT}. 
We anticipate that PACE is analogous to classical PCA, up to restricting to the largest observations (in terms of the $L^2$ norm) and considering angular components. 
% \begin{algorithm}\label{algo:PACE}
%   \caption{\textsc{PACE}: PCA of the Angular Component of Extremes}
% % \noindent\fbox{%
% %     \parbox{\textwidth}{%
% \begin{itemize}
% \item  INPUT: $X_1,\ldots, X_n$ regularly varying data ; $p<d$ a target reduced dimension.  
% \item Choose $k\ll n$ the number of observations considered as extremes. 
% \item Sort the input by decreasing value of the $L^2$ norm. The sorted sample is denoted $X_{(1)}, \ldots, X_{(n)}$. Retain the angular components associated to the largest observations
%   $$
% \Thet_{(1)}, \ldots \Theta_{(k)}, 
% $$
% where
% $\Theta_{(i)} = X_{(i)}/\|X_{(i)}\|$. 
% \item Form the second moments matrix of the above angular sample,
%   \begin{equation}
%     \label{eq:tpdmk}
%     \hat \TPDM_k = \frac{1}{k}\sum_{i=1}^k \Theta_{(i)}\Theta_{(i)}^{\T}.
%   \end{equation}
% \item Compute the eigendecomposition of $\hat\Sigma_k$.
% \item OUTPUT: Eigenvectors $\hat u_1,\ldots, \hat u_D$, sorted by decreasing values of the  eigenvalues $\hat \lambda_1\ldots, \hat \lambda_D$ ;
%   $\hat V_k = \Span(\hat u_1,\ldots, \hat u_p)$. 
% \end{itemize}

% %%}}
% \end{algorithm}

\paragraph{Structure and Organization}
\S \ref{sec:backgroundPCARV} develops extremal PCA by combining ideas from classical PCA and the probabilistic framework of regular variation.
\S \ref{sec:PCA} gives a formal, projection-based development of standard PCA, and \S \ref{sec:regvar} gives the basics of regular variation.
\S \ref{sec:PCAandRegVar} then presents the PACE algorithm. % {\em PACE:}  PCA of the Angular Components of Extremes which is analogous to classical PCA, up to restricting to the largest observations (in terms of norm) and considering angular components.
\S \ref{sec:probAndEst} presents results from \cite{DS}.
\S \ref{sec:probaSetting} discusses the properties of the TPDM and its subasymptotic versions $\Sigma_u$ and gives interpretations of the  eigenspaces of $\TPDM$ in the context of multivariate extremes; and the main   statistical guarantees derived in~\cite{DS} are presented in \S \ref{sec:estimation}.
%We then turn our attention to \cite{CT}, describing it via the already-devloped framework 
\S \ref{sec:extPCs} presents ideas from \cite{CT}, defining and discussing properties of the extremal principal components and introducing a notion of scale of a regularly varying random vector's elements.
The section also explores the $\tau$-transform's ability to model regular variation restricted to the positive orthant.
\S \ref{sec:applis} illustrates exploratory analyses performed by PACE on two data sets:  a financial data set consisting of 30 instruments and a precipitation data set for 8510 locations. \S \ref{sec:perspectives} closes this chapter with several  future research perspectives. %\as{remove functinal pca from 'perspectives' or amend the text} 

% provide the necessary background regarding PCA, typical interpretations in  applications and theoretical guarantees in a classical setting. In the same section we recall standard  Regular Variation assumptions and their immediate consequences. In \S \ref{sec:tpdm} we discuss the properties of the TPDM and its subasymptotic versions $\Sigma_u$. We give interpretations of the  eigenspaces of $\TPDM$ in the context of EVA and we introduce the  $\tau$-mapping allowing to handle positivity constraints. Finally the main   statistical guarantees derived in~\cite{DS} are presented. \S \ref{sec:applis} illustrates exploratory analyses performed by PACE on two data sets:  a financial data set consisting of 30 instruments and a precipitation data set for 8510 locations. \S \ref{sec:perspectives} closes this chapter with several  research perspectives. 

% \as{todo: more citations in this introduction? this may be a bit self centered at this point. Also is this really a good idea to start with a comparison between \cite{DS} and \cite{CT} like it is done here? it is certainly very natural to us, but maybe not for the reader. }
% % both % and namely (cite Dan's works) on the one hand, and the separate paper

%%% Local Variables:
%%% mode: latex
%%% TeX-master: "template"
%%% End:

\vspace{-0.6cm}

%% file: backgroundPCA-RV.tex
\subsection{Classical PCA as Optimal Orthogonal Projection}\label{sec:PCA}
\paragraph{Optimal Orthogonal Projection}
% \DC{Cut or include above?}
% PCA is an extremely well documented topic, and the proofs of the standard results presented  below, as well as many illustrative examples and interpretations  may be found in several textbooks, \emph{e.g.}~\cite{seber2009multivariate,joliffe} or~\cite{saporta2006probabilites}, see also the recent review paper~\cite{greenacre2022principal}. 

While the brief PCA presentation above may be adequate for practitioners, we give a more formal presentation here which will provide the foundation for the extremes-specific method presented in \S \ref{sec:PCAandRegVar}.
Proofs of the standard results presented  below, as well as many illustrative examples and interpretations,  may be found in textbooks \cite[e.g.,][]{joliffe,saporta2006probabilites, seber2009multivariate}.

To perform dimension reduction, it is possible to consider the eigendecomposition of the second moments matrix  $\mathbf{S}= \EE(\mathbf{X}\mathbf{X}\T)$ instead of the covariance matrix.
%with similar interpretation of principal axes and components, although the quantity that is being maximized by principal axes is then the second moment, not the variance, of the projections $a_j\T X$.  
In practice, this would amount to skipping the centering step and working with a data matrix whose columns have \emph{not} been centered. 
%The resulting decomposition typically features a first eigenvector which may be interpreted as an overall scaling effect,  capturing a combination of the variance in the data and the influence of the mean of the variables. 
In this case, the first eigenvector captures a combination of the variance and the mean offset of the uncentered data, and thus tends to align in the direction of the overall mean vector.
%In this case, the eigenvalues and eigenvectors of the centered covariance matrix, on the one hand, and the non-centered covariance matrix, on the other hand, are similar.  
See \cite{cadima2009relationships} for  theoretical results and further properties of uncentered PCA. 
Let $\lambda_1\ge \cdots \ge \lambda_D$ be the eigenvalues of $\mathbf{S}$, 
and $\mathbf{a}_1, \ldots, \mathbf{a}_D$ be a family of eigenvectors associated to the above eigenvalues. 

% Let $X$ be a random vector in $\rset^D$ with $\EE[\|X\|^2]<\infty$. Let $S = \EE[X X\T]$ denote the second moments matrix. 
% % $$
% % S= \EE[XX\T]
% % $$
% %In the classical setting, PCA consists of an eigen decomposition of $S$, possibly after some centering, but not necessarily so. 
% %In this chapter for simplicity we do not consider any preliminary centering step. 
% In standard practice, PCA is thought of as an eigendecomposition of the covariance matrix as presented above.
% In the sequel $S$ will be this more general second moments matrix.
% We assume that $X$ has a natural center at the origin, and the covariance matrix can be thought of in this more general setting by letting $X$ be random vector representing the data after centering about its mean.
% %In the sequel w
% We let $\lambda_1\ge \lambda_2\ge \dots \lambda_D$ denote the eigenvalues of $S$,  % and we assume for simplicity that all nonzero eigen values are distinct (in other word, with multiplicity one).
% %Also we denote by 
% and $u_1, \ldots, u_D$ denote a family of eigenvectors associated to the above eigenvalues. 

A more formal way to understand PCA is in terms of dimensionality reduction through orthogonal projection onto a linear subspace, with optimal reconstruction error or risk.
  Let $\Pi_V(\mathbf{X})$ be the orthogonal projection from $\mathbb{R}^D$ onto a linear subspace $V\subset \rset^D$, and define the \textit{reconstruction risk}\index{Dimension reduction in multivariate extremes!risk!reconstruction}  as
\begin{equation}
  \label{eq:pcaRisk}
  R(V) = \EE[\|  \mathbf{X} - \Pi_V(\mathbf{X}) \|^2]. 
\end{equation}
PCA achieves the optimal $d$-dimensional subspace in terms of reconstruction error $R$.
%The main result motivating PCA in this context is that the optimal subspace $V$ in terms of the reconstruction error $R$, among all possible eigenspaces of dimension $p$, is given by $V^* = \Span(u_1,\ldots, u_d)$. Unicity is granted as long as $\lambda_d> \lambda_{p+1}$. %\as{cite something from my lecture notes}
%In mathematical terms, 
Let $\mathcal{V}_d$ be the set of all linear subspaces $V\subset \rset^D$ of dimension $d\le D$, and let $ V^* = \Span(\mathbf{a}_1,\ldots, \mathbf{a}_d) $. 
%{\color{orange} If we assume that $\lambda_d>\lambda_{p+1}$ % then $V^*$ is an optimal projection subspace in the sense that  
%$$
 %\argmin_{V\in\mathcal{V}_d} R(V) = \{V^*\}.
 %$$}
%\DC{Is it correct to replace above with:} 
Then, $V^*$ is an optimal projection subspace in the sense that  
$$
V^* \in \underset{V\in\mathcal{V}_d}{\arg\min} \,R(V) ,
 $$
and $V^*$ is the  unique solution in the above problem if $\lambda_d>\lambda_{d+1}$.
In addition, the optimal risk is the remainder term
$$
R(V^*) = \sum_{j\ge d+1} \lambda_k.
$$
By the Pythagorean theorem, $\EE(\|\mathbf{X}\|^2) = R(V) + \EE(\|\Pi_V(\mathbf{X})\|^2)$, and the left-hand side $\EE(\|\mathbf{X}\|^2)$ does not depend on $V$. 
Thus, minimizing $R(V)$ is equivalent to maximizing the second moment of the projection,  $\EE(\|\Pi_V(\mathbf{X})\|^2)$; this argument lies at the core of the informal PCA presentation in \S \ref{sec:intro}.
\paragraph{Learning from Data}
  In practice, $\mathbf{S}$ is unknown and is replaced  with its empirical counterpart $\hat{\mathbf{S}} = 1/n \sum_{i=1}^n \mathbf{X}_i\mathbf{X}_i\T$  based on a random sample $\{\mathbf{X}_{i}\}_{i = 1}^n$,  % The empirical counterpart of $S$ is
 % \begin{equation*}
 %   \hat S = \frac{1}{n} \sum_{i=1}^n X_iX_i\T
 % \end{equation*}
 with eigen elements $\hat \lambda_1\ge \cdots \ge  \hat \lambda_D$ and $\hat{\mathbf{a}}_1, \ldots, \hat{\mathbf{a}}_D$. By the same argument as above, the \textit{empirical risk},\index{Dimension reduction in multivariate extremes!risk!empirical} 
 \begin{equation*}
   \hat R (V) = \frac{1}{n}\sum_{i=1}^n  \|\mathbf{X}_i - \Pi_V(\mathbf{X}_i)\|^2 
 \end{equation*}
 is minimized over $\mathcal{V}_d$ by $\hat V = \Span(\hat{\mathbf{a}}_1,\ldots, \hat{\mathbf{a}}_d)$.  
 
 Statistical guarantees regarding consistency of this approach have been the subject of a wealth of literature. 
 Major sources of inspiration for extending such statistical guarantees to the extreme value analysis context in~\cite{clemenccon2023regular,DS} %\as{to be modified}) 
are~\cite{blanchard2007statistical} and~\cite{zwald2006convergence}. In the first reference the goal is to establish non-asymptotic upper bounds on the \textit{excess risk},\index{Dimension reduction in multivariate extremes!risk!excess} 
 $$
R(\hat V) - R(V^*). 
 $$
 Upper bounds of order $O(1/\sqrt{n})$ are obtained following classical steps in the theory of empirical risk minimization; that is, using uniform bounds on the deviations $\hat R(V) - R(V)$ following \cite{shawe2002eigenspectrum,shawe2005eigenspectrum}.   Fast rates of convergence of order $O(1/n)$ are obtained using a localized approach in the non-centered case.  
 In contrast, in~\cite{zwald2006convergence}, the deviations of the empirical eigenspace $\hat V$ are controlled using perturbation theory applied to the covariance operator, which is the approach taken in~\cite{clemenccon2023regular}. %\as{here again}. 

\subsection{Regular Variation}\label{sec:regvar}\index{Regular variation!multivariate}
%Needed assumptions/opbjects objects:
%\as{it is more convenient for me to put $b(s)$ outside the probability sign, and then choose it as $\PP[\|X\|>s]^{-1}$. }
%Regular variation is a widely used minimal assumption in multivariate extreme value theory. 
%is a baseline assumption for the analysis of multivariate extremes.  
\paragraph{Multivariate Regular Variation}
At its most basic level, multivariate regular variation characterizes the tail behavior of multivariate heavy-tailed distributions. The definition given below roughly implies that the individual components of the random vector $\mathbf{X}$ all have tails like power functions with a common index $\alpha > 0$.
Formally, we assume
%it assumes that 
$\mathbf{X}$ satisfies %assumptions are
\begin{equation}
  \label{eq:regvar}
  b(u) \, \text{P}(  \mathbf{X}\in u B) \to \nu(B), 
\end{equation}
on Borel sets $B$ bounded away from the origin, and  % \as{cite resnick87} or M-convergence \as{don't forget Meerschaert, Daley Vere-Jones maybe}
where $b(u)$ is a regularly varying function, that is, $b(u) = \mathcal{L}(u) u^{\alpha}$  for some regular variation index $\alpha>0$ and some \emph{slowly varying} function $\mathcal{L}$, meaning that $\mathcal{L}(ux)/\mathcal{L}(u)\to 1$ as $u\to \infty$, for all $x>0$.  Note that unless 
 $\alpha > 2$, $\mathbf{X}$ has no finite second moments. This highlights once more why classical PCA is inadequate for extremes, as the classical literature typically requires the existence of fourth moments. 
%\DC{Shouldn't $b(s) = L(s)s^\alpha$? }\as{yes!} % Or maybe it belongs in the denominator?  If $b(s) =\PP[\|X\|>s]$, then it belongs in the denominator, right?}%\as{ notice the exponent $-1$}}

See, for example, \cite{bingham1989regular,resnick2007,resnick2008extreme} for classical introductions to regular variation, or to \cite{hult2006regular, meerschaert1984multivariable} for alternative presentations allowing extensions to general metric spaces. 
%Notice that 
The  apparently strong assumption~\eqref{eq:regvar} is in fact automatically satisfied after marginal standardization when the data belong to some multivariate maximum domain of attraction, with potentially nonstandard margins, (\cite{resnick2008extreme}, Propositions 5.10 and 5.14). 
% )\as{see Resnick 87 chapter 5.10, 5.14},
%which is a widely used minimal assumption in multivariate extreme value theory. 
Hence, regular variation is a widely-used minimal assumption in multivariate extreme value theory.

%We can and will 
%\textcolor{red}{
\paragraph{Equivalent Representations}
Over this section we follow~\cite{DS} and take
\begin{equation}\label{ch12:rfg}
  b(u) =\frac{1}{\PP(\|\mathbf{X}\|>u)}. 
\end{equation}
With this choice of a scaling function $b$, the restriction of $\nu$ to $\rset^D\setminus \ball$ is a probability measure which we denote by $P_\infty$, where  $\ball = \{\mathbf{x} \in \mathbb{R}^D : \| \mathbf{x} \| \le  1\}$ is the unit ball.

An equivalent characterization of~\eqref{eq:regvar}  is through the weak convergence of $P_u = \PP(\mathbf{X}\in u\,\cdot \mid  \|\mathbf{X}\|>u)$. 
%\DC{What is $E$?  A general cone/set in $\mathbb{R}^p$ such that if $x \in E, tx \in E$ for all $t > 0?$ Do we want to say ``often $E = \mathbb{R}^p \setminus \{0\}$ or $E = \mathbb{R}_+^p \setminus \{0\}$?"} \as{remove $E$ and wimply work in $\rset^p$}
Indeed,~\eqref{eq:regvar} is equivalent to  
\begin{equation}
  \label{eq:equivRegVarWeakcv}
P_u \wto P_\infty  \quad \text{ and } \quad  \PP(\|\mathbf{X}\| > u) = \frac{\mathcal{L}(u)}{u^{\alpha}},  
\end{equation}
%\DC{$s$ or $x$ in statement above?}
for some slowly varying function $\mathcal{L}$, where $\wto$ denotes weak convergence of probability measures as $u\to\infty$;  see \cite{segers2017polar} for a proof.
Let $\mathbf{X}_\infty\sim P_\infty$. By definition, for any measurable set $A\subset \rset^D \setminus \ball$ such that $\nu(\partial A)=0$, 
$$
\PP(\mathbf{X}_\infty\in A) = \nu(A) = \lim_{u\to \infty} \PP(\mathbf{X} \in u A \,|\, \|\mathbf{X}\|\ge u). 
$$
We use the $L_2$ norm for simplicity but other norms could be readily used. In line with the previous chapters, we define the angular vector 
$$ \boldsymbol{\Thet} = \frac{\mathbf{X}}{\| \mathbf{X} \|}, $$ %\theta(X). $$
which has elements between zero and one and satisfies $\Thet\T \Thet = 1$. 
A third  condition which is equivalent to~\eqref{eq:regvar} and~\eqref{eq:equivRegVarWeakcv} is 
\begin{equation}
  \label{eq:equivRegvarAngular}
   \mathbb{L}( \| X \| / u, \boldsymbol \Theta \, | \, \| X \| >u ) \wto \Pareto(\alpha)\otimes H. 
\end{equation}
Here, $\mathbb{L}(Z)$ denotes the law of the random vector $Z$, and $H$ is the \emph{angular measure} with respect to $\nu$,  which is defined on the unit sphere $\mathbb{S} = \{\mathbf{x} \in [0,\infty)^D : \|\mathbf{x}\|=1\}$. See also Chapter~7.\footnote{Eq.~\eqref{eq:equivRegvarAngular} is connected to the radial-angular independence property discussed in earlier chapters.} %; see Chapter~\ref{ch:regmulti}.} 

\paragraph{Takeaway}
With the standardization in \eqref{ch12:rfg}, $H$ is a probability measure and we denote by $\boldsymbol \Theta_\infty$ a random vector distributed according to $H$ on $\sphere$. %  Similarly let $R_\infty\sim\Pareto(\alpha)$ be independent from $\boldsymbol \Theta_\infty$ 
The angular measure $H$ contains all dependence information in the limiting measure $\nu$.
In high dimensions, $H$ is difficult to fully characterize, motivating the need for PCA.

%%%%%%%%%%%%%%%%%%%%%%%%%%%%%%%%%%%%%%%%%%%%%%%%%%%%%%%%%%%%%%%%%%%%%%%%%%%%%%%%%
%%%%%%%%%%%%%%%%%%%%%%%%%%%%%%%%%%%%%%%%%%%%%%%%%%%%%%%%%%%%%%%%%%%%%%%%%%%%%%%%%

\subsection{PCA for Regular Variation}\label{sec:PCAandRegVar}

%The general setting and motivation behind both works are the same:  
To develop PCA for extremes, both \cite{CT} and \cite{DS} consider a $D$-dimensional  random vector $\mathbf{X}$ satisfying the regular variation assumption (\ref{eq:regvar}). % (See Section~\ref{sec:regvar}). %, that is, no standardization is required for a limit measure $\nu$ to exist.
%We denote by $\alpha>0$ the regular variation index.
The assumption underlying a PCA approach is that  when $d$ is large, the
  support of the limit measure $\nu$  is concentrated on (or in the vicinity of) a lower dimensional
  subspace of $\rset^D$.
This means that some particular linear combinations of the components
  are much more likely to be large than others. 
Identifying this subspace
  and thus reducing the dimension would facilitate a refined
  statistical analysis.
 In practice, often one  fixes a target reduced dimension $d<D$, and then aims to build a $d$-dimensional representation of $\mathbf{X}$ capturing as much information as possible.

To adapt the classical PCA methodology to the analysis of extreme events, consider  the (non-centered) covariance matrix 
of  extreme angles above threshold $u$, %of observations which norm exceeds the threshold $t>0$, %and their eigendecomposition. To fix ideas, consider the
\begin{equation}
  \label{eq:TPDMt}
  \Sigma_u = \EE(\boldsymbol{\Thet}\boldsymbol{\Thet}\T \,|\, \|\mathbf{X}\|>u).  
\end{equation}
% this key covariance matix is defined as
%where $\Thet \X / \| \X \|$ is the angular component of the random vector $\X$. %, namely $\Thet = \X / \| \X \|$.
In~\cite{DS}, a slightly more general definition is given for $\Thet$, which is defined as any  rescaled version of $\mathbf{X},$ $\omega(\mathbf{X})\mathbf{X}$ where the scaling function $\omega$ is chosen so that $\omega(\mathbf{X})\mathbf{X}$ has enough moments. 
In this chapter, we trade generality for simplicity and we limit ourselves to the case  where $\omega(\mathbf{x}) = 1/\|\mathbf{x}\|$.
%\DC{Additional text commented out here.  Anne:  please check to make sure it is okay to cut.}\as{ok} %a bit sad not to mention the lead to analyse the centering step in Blanchard07. willthink. }
%which ensures compatibility with~\cite{CT}'s framework. The choice of disregarding any centering pre-processing step is also driven by simplicity in~\cite{DS}; refined analysis could likely allow to handle such empirical centering without altering the nature of the theoretical guarantees obtained so far, as it is the case in the classical setting, see  \emph{e.g.}~\cite{blanchard2007statistical}. 
%\as{here say a word about centering/not centering,}

Under classical  regular variation assumptions further detailed in \S \ref{sec:probaSetting},  the limit as $u\to \infty$  in~\eqref{eq:TPDMt}  exists, and is referred to as  the \emph{Tail Pairwise Dependence Matrix}\index{Tail pairwise dependence matrix} (TPDM) in \cite{CT}, which we denote here  by $\TPDM$, 
\begin{equation}
  \label{eq:TPDM}
  \TPDM = \lim_{u\to \infty}\EE(\boldsymbol{\Thet}\boldsymbol{\Thet}\T \,|\, \|\mathbf{X}\|>u) = \EE(\boldsymbol \Theta_\infty \boldsymbol \Theta_\infty\T). 
\end{equation}
%As shall become clear in Section
%The covariance matrix for extreme angles $\TPDM$ is called the \emph{Tail Pairwise Dependence Matrix} (TPDM) in \cite{CT}.
In~\cite{DS}, the limit $\TPDM$ is not explicitly introduced and the focus is on $d$-dimensional subspaces  $V$ of $\rset^D$ such that the orthogonal projection of $\mathbf{X}$ onto $V$  minimizes a reconstruction error analogous to (\ref{eq:pcaRisk}), but in terms of the angular components (see \S \ref{sec:probaSetting}).

The PACE algorithm introduced below can be seen as an analogue to classical PCA, but applied to the {\em angular component}\index{Pseudo-polar representation!angular component} $\boldsymbol \Theta$ of a regular varying random vector.  
As with PCA, the algorithm consists of an eigendecomposition, with an estimate of the TPDM replacing the role of the estimated covariance matrix.
Further justification of the algorithm follows in \S \ref{sec:probAndEst}.
% \vspace{-0.1cm}
% \end{shortbox} \vspace{1cm}
\begin{algorithm}
  \caption{\textsc{PACE}: PCA of the Angular Component of Extremes\label{algo:PACE}}
% \noindent\fbox{%
%     \parbox{\textwidth}{%
  \textbf{Input}: $\mathbf{X}_1,\ldots, \mathbf{X}_n$ regularly varying data; a target reduced dimension $d<D$.
\begin{enumerate} 
\item Choose $k\ll n$ the number of observations considered as extremes. 
\item Sort the input by decreasing value of the $L^2$ norm, $\|\mathbf{X}\|$. Retain the angular components associated to the largest $k$ observations
  $$
\{\boldsymbol \Theta_{1}, \ldots, \boldsymbol \Theta_{k}\} = \left\{\frac{\mathbf{X}_{i}}{\|\mathbf{X}_{i}\|}: \|\mathbf{X}_{i}\| > u \right\}. 
$$
\item Form the second moments matrix of the above angular sample,
  \begin{equation}
    \label{eq:tpdmk}
    \hat \TPDM_k = \frac{1}{k}\sum_{i=1}^k \boldsymbol \Theta_{i}\boldsymbol \Theta_{i}\T.
  \end{equation}
\item Compute the eigendecomposition of $\hat\Sigma_k$.
\end{enumerate}
\textbf{Output}: Eigenvectors $\hat{\mathbf{a}}_1,\ldots, \hat{\mathbf{a}}_D$, sorted by decreasing values of the  eigenvalues $\hat \lambda_1,\ldots, \hat \lambda_D$; empirical eigenspace 
  $\hat V_k = \Span(\hat{\mathbf{a}}_1,\ldots, \hat{\mathbf{a}}_d)$. 
%%}}
\end{algorithm}

%% file: probaPCAExtremes.tex
In this section, we investigate the large-sample properties of the eigendecomposition of $\Sigma_u$ in (\ref{eq:TPDMt}), recalling that the true target is the eigendecomposition of the TPDM $\Sigma$ in (\ref{eq:TPDM}). We also make a connection between principal eigenspaces (i.e., $\mbox{Span}(\mathbf{a}_1, \ldots, \mathbf{a}_d) \mbox{ for } d \in {1, \ldots, D}$) of $\TPDM$ and the support of the limit measure $\nu$ (or equivalently, $P_\infty$ or $H$).  
%\DC{By eigenspaces, we mean the span of $u_1, \ldots, u_d$, right?  The definition given here:  https://mathworld.wolfram.com/Eigenspace.html talks about the space created by the eigenvectors associated with a single eigenvalue.  Do you think we could define ``eigenspace" as ``the space spanned by the first $p$ eigenvectors."  ``Eigenspace appears many times in the following text.} 
\paragraph{Warm-up: Existence of TPDM and its Consequences}
As a warm-up, we first establish  the existence of the TPDM matrix  $\TPDM$ defined in~\eqref{eq:TPDM}. % the second moment matrix of the angular component at infinity, $\boldsymbol\Theta_\infty$, introduced in \S \ref{sec:regvar}.
% Recall Definition~\eqref{eq:TPDM} fro $\TPDM$ and n
%{\color{orange} Beginning with $x \in \mathbb{R}^D$, let $r = \|x\|_2$ and $\theta = x/r$.
%As the mapping $(r,\theta)\mapsto \theta\theta\T$} is bounded and  continuous on $\rset_+\setminus\{0\}\times\sphere$. 
%\DC{Although we've talked about polar coordinates earlier in the regular varying context, I don't think we've defined $(r, \theta)$ previously.  Maybe we just need to say $(r, \theta) \in [0, \infty) \times [0, 2\pi]$?  And maybe say "polar representation"?}
%\as{I just commented the $r,\theta$ discussion because looking at \eqref{eq:equivRegvarAngular} we already have weak cv in polar coordinates (it's not needed to mention that marginalisation is a ocntinuous mapping, this is very basic), but below is a slight modification} 
Because the mapping $\thetab\mapsto \thetab\thetab\T$ defined on $\mathbb{S}$ is bounded and continuous, 
the weak convergence characterization in \eqref{eq:equivRegvarAngular} implies immediately that
$\EE(\boldsymbol\Theta\boldsymbol\Theta\T \,|\, R>u) \to \EE(\boldsymbol\Theta_\infty\boldsymbol\Theta_\infty\T)$ as $u \to \infty$. Thus, the limit in~\eqref{eq:TPDM} indeed exists and  is given by % we have
\begin{equation}
  \label{eq:identityTPDM}
  \TPDM = \EE(\boldsymbol\Theta_\infty\boldsymbol\Theta_\infty\T). 
\end{equation}
In words, $\TPDM$ exists and is the second moment matrix of $\boldsymbol\Theta_\infty$. In this respect, $\widehat \Sigma_k$ in Algorithm~\ref{algo:PACE} may be interpreted as an estimator of $\Sigma$.

\paragraph{Back to Risk Minimization}
We follow the risk minimization view on PCA introduced in \S \ref{sec:PCA} and apply it to $\boldsymbol\Theta_\infty$, which encapsulates the dependence structure of extremes.
Importantly, the angular component $\boldsymbol\Theta_\infty$ is bounded and thus has finite moments of all orders thereby meeting the fourth moment requirement for consistency results of PCA estimators.
%has the double advantage of $(i)$ encapsulating the dependence structure of extremes; $(ii)$ being bounded, and as such, having finite moments of all orders. 
%In this spirit it is natural to introduce 
We define the reconstruction risk $R_\infty$ for $\boldsymbol\Theta_\infty$ relative to a projection subspace $V$ defined as 
   \begin{equation}\label{eqn:Rinfinity}
    R_\infty(V) = \EE[ \| \Thet_\infty - \proj_V(\boldsymbol{\Thet}_\infty)\|^2  ]. 
  \end{equation}
As noted in \S \ref{sec:PCAandRegVar}, extremal PCA is most useful when the limit measure $\nu$ lies in the vicinity of a lower dimensional subspace of $\mathbb{R}^D$; that is, a subspace $V$ can be found such that $R_\infty(V)$ is small.
%Similar to traditional PCA and as stated earlier,  framework is particularly useful  when the data concentrate in the vicinity of a lower dimensional subspace, so that the associated reconstruction risk is small. 
%Similarly,  
%extremal PCA is useful when the {\em large}  
%likely to produce interpretable results when the limit measure $
The assumption in Proposition~\ref{prop:concentrationMu} below  is stronger than $\nu$ being in the vicinity of a lower dimensional subspace and may be interpreted as a limiting case of this setting. 
Saying $\nu$ is concentrated on a subspace $V_\nu\subsetneq\rset^D$, implies $\nu(\rset^D\setminus V_\nu)=0$.
Unlike similar results in classical PCA, which assume that the support of $\mathbf{X}$ is constrained to this subset, the regular variation framework implies that the support of $\mathbf{X}$ {\em as it grows large} becomes limited to $V_\nu$.
This can be made precise in the following two ways.
%Considering such a setting is not new in EVA and is central in the  hidden regular variation framework (see \cite{das2013living} and the references therein). 
%The tail measure concentrating on a subspace means roughly that certain linear combinations of components "cannot" be large, more precisely that the probability that they are large is negligible compared with the probability that one of the components is  large.  
%Mathematically, 
If $\nu$ concentrates on $V_\nu \subsetneq\rset^D$ and if $\textbf{a}\in (V_\nu)^\perp\setminus\{0\}$ (the orthogonal complement of $V_\nu$), then for any norm $\|\cdot\|$, including the infinite norm, for any $\varepsilon>0$, as $u \to \infty$,
$$
\frac{\PP(\textbf{a}\T \textbf{X} / u> \varepsilon )}{\PP(\|\mathbf{X}\|>u)  }\to 0.
$$
%Another simple interpretation is the following: concentration of $\nu$ on $V_\nu$ means that   tail events concentrate in the neighborhood of $V_\nu$. Mathematically, 
Alternatively, consider the distance between $\textbf{x}\in\rset^D$ and a set, i.e., the \textit{point-to-set distance}, $d_{\text{ps}}(\textbf{x},V)  = \inf\{\|\textbf{x}-\textbf{y}\|: \textbf{y} \in V\}.$ 
Then, under the assumption that $\nu$ concentrates on $V_\nu$, we have that as $u \to \infty$, 
$$
\PP\{d_{\text{ps}}(\textbf{X}/u,V_\nu) < \varepsilon \,| \, \|\mathbf{X}\|>u \}\to 1. 
$$
Both  interpretations complement each other and offer an alternative notion of sparsity compared with existing ones related to  multiple subspace identification  and clustering; see, e.g., \cite{butsch2024information,chiapino2016feature,chiapino2019identifying,goix2016sparse,goix2017sparse,janssen2020k,meyer2021sparse}, to name a few. 
%where the definition of sparsity for extremes relies on the idea that "only certain subgroups of components may be simultaneously large". }

%  The next claim  is an immediate consequence of Lemma 2.5 in~\cite{DS}.
%\as{todo make a clearer statement below, a word about sparse support of extremes}  
%\begin{shadebox}\vspace{-0.4cm}
\begin{proposition}\label{prop:concentrationMu}
  If $\nu$ is concentrated on a $d^*$-dimensional subspace $V_\nu\subset \rset^D$  and if no   subspace $\tilde V\subset V_\nu$ of dimension strictly less than $d^*$ contains the support of $\nu$, then:
  \begin{enumerate}
  \item $\argmin_{V\in\mathcal{V}_{d^*}} R_\infty(V) = \{V_\nu\} $ and  $R_\infty(V_\nu) = 0$. %, and. 
  \item $\TPDM$ has  exactly $d^*$ nonzero eigenvalues (with multiplicities).
  \item  $V_\nu = \Span(\textbf{a}_1,\ldots, \textbf{a}_{d^*})$. 
  \end{enumerate} 
\end{proposition} %\vspace{-0.4cm}
%\end{shadebox}
% \as{todo: proof (see \cite{DS}, Lemma 2.5)}
%\DC{$d^*$ the (true) dimension of the subspace $V_{\nu}$, correct?  Should we say this?  I don't think $d^*$ has appeared previously.}\as{correct, and your orange addition in the statement does the job}
\noindent Proposition~\ref{prop:concentrationMu} is an immediate consequence of Lemma 2.5 in~\cite{DS}.

\paragraph{A More Realistic Setup}
In practice, $\nu$ is not likely to be concentrated on a subspace $V_\nu$ such as the one in Proposition~\ref{prop:concentrationMu}, and even so,  the dimension $d^* $ would typically be unknown.  %however concentration on a neighborhood of a subspace would result in a spectral gap  which could be detected on a screeplot.
%The choice of $p$ is a typical issue in classical PCA, with no off-the-shelf solution. 
As it is with classical PCA, there is no general theoretical answer
to the question of how to choose $d$ in the extreme-value context. %and this limitation passes onto the EVA setting. %As a consequence here we fix $p$
%In practice the choice of the reduced dimension $p$ depends on the downstream task: If the goal is simply to reach a reasonable compromise between complexity and accuracy, the choice of $p$ may be done \emph{via} a scree-plot.  
In the context of developing probabilistic results, we fix the reduced dimension $d < D$ and the goal pursued is to solve the minimization problem %estimate  $$V_\infty= \argmin_{V\in\mathcal{V}_d} R_\infty.$$
\begin{equation}
  \label{eq:riskminiproblem}
  V_\infty = \arg \min_{V\in\mathcal{V}_d} R_\infty (V),   
\end{equation}
based on an empirical version of $R_\infty$ in \eqref{eqn:Rinfinity}.\footnote{The classical arguments presented in \S \ref{sec:PCA} and applied to $\boldsymbol\Theta_\infty$ ensure again that the solution of the minimization problem in \eqref{eq:riskminiproblem} is unique as long as $\lambda_d>\lambda_{d+1}$, which we assume henceforth. In this context $V_\infty$ is the unique principal eigenspace of $\TPDM$ of dimension $d$.}

The  quality of an approximation  subspace $W$ for $V_\infty$ is measured in \cite{DS} in terms of an appropriate notion of distance $d$, defined as  the operator norm between orthogonal projectors onto linear subspaces,
$$
d(V,W) = \sup_{\mathbf{x}\in\sphere}\|\proj_V \mathbf{x} - \proj_W \mathbf{x}\|. 
$$

%   The goal of PCA  (for fixed $p$)  can be restated as
  
% \noindent\fbox{%
%   \parbox{\textwidth}{%
%     Find $\widehat V $ such that $R_\infty(\widehat V) - R_\infty(V_\infty)$ is small, or at least such that
%     $R_u(\widehat V) - R_u(V_u)$ for larrge $t$, is small, and such that
%     $$
% d(\widehat V, V_\infty)
% $$
% vanishes for large sample sizes, where $d$ is a measure of the distance between subspaces. 
% }}

As it is generally the case in extreme value theory,  the error induced by  considering  $\widehat \TPDM_k$ in Algorithm~\ref{algo:PACE} rather than the unknown matrix $\TPDM$ results from two sources.
First the asymptotic matrix $\TPDM$ is approximated by the subasymptotic version $\Sigma_u$ in~\eqref{eq:TPDMt} where $u$ is chosen  as the quantile  of order  $1-k/n$ for the norm  $\|\mathbf{X}\|$, which we denote by $u_{n,k}$. Second,  $\TPDM_{u_{n,k}}$ itself is approximated by the empirical matrix $\widehat \Sigma_k$. In this section, we analyze the first source of error (a bias term), while the second source of error (a variance term) is considered in \S \ref{sec:estimation}.  %which is the $k^{th}$ larger order statistic of the norm ; Second, $\Sigma

% The analysis carried out in~\cite{DS} decomposes  
%With this program in mind, %\as{rewriting to be continued below}
% \begin{itemize}
% \item  We 
  Consider the \textit{conditional reconstruction risk} above level $u$,\index{Dimension reduction in multivariate extremes!risk!conditional reconstruction}
  \begin{equation}\label{eq:subasymptoticRisk}
    R_u(V) = \EE(\| \Thet - \proj_V(\Thet)\|^2 \,|\, \|\mathbf{X}\|>u), 
  \end{equation}
  % and for the limit distribution
 and denote by $V_{u}$ a minimizer of $R_u$, which is also a principal eigenspace of $\TPDM_u$ of dimension $p$ % , which  is also $ \argmin_{\mathcal{V}_d} R_u(V)$
  (Lemma 2.1 in \cite{DS}).
%\
  
  By the weak convergence property~\eqref{eq:equivRegvarAngular}, it holds that for all fixed linear subspace $V$,   $R_\infty(V) = \lim_{t \rightarrow \infty} R_u(V)$ (Proposition 2.2 in~\cite{DS}). % However this is not sufficient to ensure convergence of the excess risk $R_\infty(V_u) - R_\infty(V_\infty)$ since $V_u$ varies with $t$.
   The  main probabilistic result  in~\cite{DS} is the convergence of the risk minimizer (Theorem 2.4 in \cite{DS}):
%\begin{shadebox}\vspace{-0.4cm}
   \begin{proposition}[Convergence of eigenspaces of $\Sigma_u$]\label{prop:mainProbaResult} 
    As $u\to\infty$, $$d(V_u,V_\infty)\to 0. $$
  \end{proposition}% \vspace{-0.4cm}
% \end{shadebox}  
\noindent This result  follows from standard arguments % As it is generally the case in the present framework of
for contrast minimisation (or M-estimation, or risk minimisation, depending on the authors).  Namely the proof relies on   % , the pointwise convergence is not sufficient to guarantee convergence of the minimizers. However, following standard arguments,   it is possible to do so using  Lipschitz continuity properties of $R_u$  are required.  In \cite{DS}
  % Following standard arguments in  and based on
 uniform Lipschitz continuity of  $R_u$, in combination  with compactness of $\mathcal{V}_d$ with respect to $d$. %, from which the results follows. 

  % \begin{equation}
  %   \label{eq:mainProbaResult}
  %   \text{As } t\to\infty, \quad d(V_u,V_\infty)\to 0. 
  % \end{equation}
  
%\item

% For the time being we leave the estimation problem  aside (this is the subject of \S \ref{sec:estimation}) and we investigate the consistency of the eigen decomposition of $\Sigma_u$, that is the PCA of the subasymptotic version of $\TPDM$. 
% %\item

  % Probabilistic results in \cite{DS}:
  % \begin{itemize}
  % \item (Proposition 2.2):
  %   $\lim R_u(V) = R_\infty(V)$, for all $V$.
  % \item By standard arguments, this yields the following main probabilistic result
  %   (Theorem 2.4): $\lim_ud(V_u, V^*) = 0$. 
  % \end{itemize}

%\end{itemize}

%%% Local Variables:
%%% mode: latex
%%% TeX-master: "template"
%%% End:

%% file: statsDS.tex
We now turn our attention to the statistical error. Let $u \equiv u(n,k)$ be the $1-k/n$ quantile of the norm $\|\mathbf{X}\|$. The empirical counterpart of $R_u$ in \eqref{eq:subasymptoticRisk} is
    \begin{equation*}
      \hat R_k (V) = \frac{1}{k} \sum_{i=1}^k \|\boldsymbol\Theta_{i} - \proj_V(\boldsymbol\Theta_{i}) \|^2.
    \end{equation*}
    Again, note that for fixed dimension $d$ the principal eigenspace of $\hat\Sigma_u$ is the minimizer of $\hat R_u$.

The purpose of~\cite{DS}'s statistical analysis is to control  the excess of subasymptotic risk $R_{u(n,k)}$ attached to  the empirical minimizer $\hat V_k$. %, which is also a principal eigenspace of $\hat\Sigma_k$.  % where $u(n,k)$ denotes here the $1-k/n$ quantile of the norm $\|X\|$.
Thus, their  theoretical analysis  focuses on the reconstruction error attached to  $\hat V_k$ rather than on the convergence of $\hat \TPDM_k$ itself. 
 The first type of statistical results obtained in ~\cite{DS} are consistency results. For any $V$, as $k,n\to\infty$ while $k/n\to 0$, if follows that \cite[Proposition 2.6]{DS}, 
    \begin{equation*}
      \hat R_u(V) \pto R_\infty(V). 
    \end{equation*}
    Using arguments pertaining to M-estimation theory, the authors also obtain the consistency of the principal eigenspaces \cite[Theorem 2.7]{DS}.
%    \begin{shadebox}\vspace{-0.4cm}
     \begin{proposition}If $k,n\to \infty$ while $k/n\to 0$, then %under the same assumptions regarding $k,n$,  
    \begin{equation*}
       d(\hat V_k, V_\infty) \pto 0.
    \end{equation*}    
  \end{proposition}%\vspace{-0.4cm}
 % \end{shadebox}     
%\DC{Does it make sense to make this a proposition like Proposition 2 above?}\as{yes, will do}
\noindent The second type  of statistical  results in~\cite{DS} are uniform bounds on the deviation of $\hat R_k$ \cite[Theorem 3.1]{DS}.  %. %; of concern the excess risk  for $\hat V_k$,
%  \begin{shadebox}\vspace{-0.4cm}
    \begin{proposition}[Upper bound on the reconstruction risk]\label{prop:upperB_reconstruction}
      With probability at least $1-\delta,$ with $\delta \in (0, 1)$, 
   \begin{equation*}
     \begin{aligned}
       \sup_{V\in\mathcal{V}_d} | \hat R_k(V) -  R_{u(n, k)}(V) |
       & \le
         \sqrt{\frac{\min(d,D-d) (1-k/n)\tr(\Sigma_{u(n,k)}^2) }{k} }   \\
       & \hspace{0.2cm} + \sqrt{\frac{ 8(1+k/n)  \log(4/\delta)}{k}}  +  %\dots \\
        \frac{ 4 \log(1/\delta)}{3k}.  
     \end{aligned}
         \end{equation*}
       \end{proposition}% \vspace{-0.4cm}
     % \end{shadebox}
%\DC{Also make this a proposition?}  \as{yes, will do}
\noindent Some technical comments are in order:
     \begin{itemize}
     \item   Although the quantity $\emph{tr}(\Sigma_{u(n,k)}^2)$ is unknown,  it is bounded above by one. Indeed it equals the  sum of the eigenvalues of $\Sigma_{u(n,k)}^2$,  that is $\sum_{j=1}^D \lambda_{u(n,k),j}^2$ where the $\lambda_{u(n,k),j}$ are the eigenvalues of $\Sigma_{u(n,k)}$. Since $\|\boldsymbol\Theta\|\le 1$, we obtain  $\sum_{j=1}^D \lambda_{u(n,k),j} = 1$, and the result follows.  Also  Theorem~3.3 in~\cite{DS} provides a refined result with a data-dependent term instead of $\Sigma_{u(n,k)}$ which may be significantly less than one. 
     \item The upper bound above is a minimal guarantee (as opposed to a tight one), since it is to be expected that tighter bounds with a leading term of order $O(1/k)$ may be obtained at the price of further technicalities, by localization techniques as in~\cite{blanchard2007statistical}.
     \item From these uniform bounds, one may immediately derive bounds on the excess risk of $\hat V_k$ (\cite[p.~918]{DS}), since
\begin{align*}
  R_{u(n,k)}(\hat V_k) - R_{u(n,k)}(V_u)\le 2 \sup_{V\in\mathcal{V}_p} | \hat R_k(V) -  R_{u(n,k)}(V) |. 
\end{align*}\vspace{-0.4cm}
     \end{itemize}
    
    A simulation study is performed in~\cite{DS}, illustrating the potential benefits of an extremal PCA dimension reduction step before estimating the angular measure $H$.

%% file: cooleyThibaud.tex
A focus of \cite{CT} is the construction of the principal components and interpreting them via the lens provided by the eigenbasis. 
Much of \cite{CT} is motivated in the specific case of nonnegative data.
Below, we define and state properties of the principal components within the general framework described above.
We then will discuss \cite{CT}'s approach for modeling in the positive orthant via the $\tau$-transform.

%%%%%%%%%%%%%%%%%%%%%%%%%%%%%%%%%%%%%%%%%%%%%%%%%%%%%%%%%%%%%%%%%%%%

\subsection{Extremal Principal Components}\index{Dimension reduction in multivariate extremes!extremal PCA (Principal Component Analysis)} 
\vspace{-0.5cm} \strut \hfill \textcolor{gray}{\footnotesize \cite{CT, DS, kiriliouk2022}}
\paragraph{Extremal Principal Components are Based on TPDM}
In the same way that standard principal component analysis arises from the eigendecomposition of the variance-covariance matrix, extremal principal components arise from the eigendecomposition of the TPDM.

To make this precise, we follow \cite{CT} and consider any function $b(t)$ for which (\ref{eq:regvar}) holds. This implies that the TPDM $\Sigma = [\sigma_{i,j}]_{i,j = 1, \ldots, D}$ is
\begin{equation}
  \label{eq:TPDMCT}
  \sigma_{i,j} = \int_{\sphere} w_i w_{j} \,\text{d}H( \mathbf{w}).
\end{equation}
This differs from (\ref{eq:TPDM}) only in relaxing the assumption \eqref{eq:regvar}, and so $H$ is not assumed to be a probability measure.\footnote{Different choices for $b(t)$ in (\ref{eq:regvar}) result in different total mass for $H$ \cite[p. 174]{resnick2007}.} Following \cite{CT}, below we consider the simplest canonical regularly varying function, $b(u) = u^\alpha$.

An advantage of defining $\sphere$ via the $L_2$ norm is that the TPDM's diagonal elements give the total mass of $H$, 
\begin{equation}
  \label{eq:totalMass}
  \sum_{j = 1}^D \sigma_{j,j} = \int_\sphere \sum_{j = 1}^D w_j^2 dH(\mathbf{w}) = \int_\sphere \,\text{d}H(\mathbf{w}), 
\end{equation}
given that $\textbf{w} = (w_1, \dots, w_D)\T \in \mathbb{S}$. 
The elements of the regularly varying random vector $\mathbf{X}$ all have a common tail index $\alpha$, but their relative magnitudes can be compared by a notion of scale.
%\cite{CT} discuss a notion of scale for the elements of a regularly varying random vector.
For a given normalizing function $b(u)$, 
%a notion of the scale of the elements of regularly varying random vector $X$ is given by 
\begin{equation}\label{eq:scale}
%\lim_{t \rightarrow \infty} b(t)\PP(X_j > tx) = c_jx^{-\alpha},
\lim_{u \rightarrow \infty} b(u)\PP(|\mathbf{X}_j| > u) = c_j,
\end{equation} 
where $c_j = \int_\sphere |w_j|^\alpha \,\text{d}H(\mathbf{w})$.
Much like in classical PCA where the scale (i.e., variance) of the random variables is given by the diagonal elements of the covariance matrix, if $\alpha = 2$ then $c_j = \sigma_{j,j}$, and the diagonal elements of the TPDM contain these scales.
If combined with (\ref{eq:totalMass}), then the sum of these scales equals the total mass $\int_\sphere \,\text{d}H(\mathbf{w})$.
Furthermore, as $\sum_{j = 1}^D \sigma_{j,j} = \sum_{j = 1}^D \lambda_j$, when $\alpha = 2$ the eigenvalues have the familiar interpretation where $\lambda_i/\sum_{j = 1}^D \lambda_j$ is the proportion of total scale explained by the $i$th eigenvector.
Because of these convenient connections to classical PCA, \cite{CT} assume $\alpha = 2$, 
%and in the 
%\cite{CT} assume that $\alpha = 2$ which implies $c_j = \sigma_{j,j}$, thereby lending a scale interpretation to the diagonal elements of the TPDM.
%applications studied in \cite{CT}, 
and marginal transformations were applied in their applications so that this assumption could be made.
In this chapter, we will not assume $\alpha = 2$ unless explicitly stated.
Recently, \cite{kiriliouk2022} defined a tail dependence matrix with elements  $\sigma_{j,j'} = \int_{\sphere} \theta_j^{\alpha/2}  \theta_{j'}^{\alpha/2} dH( \theta)$ which preserves the aforementioned scale relationship with the diagonal elements for general $\alpha$, but we will restrict attention to the formulation given in (\ref{eq:identityTPDM}) or (\ref{eq:TPDMCT}).
%{\color{orange} In classical PCA, the scale of the random variables is described by each element's variance, given by the diagonal elements of covariance matrix $S$.}

% \DC{Anne:  I've added the paragraphs below to address a reviewer comment.  Please look over carefully.  Thanks!}\as{done, see my sugggestion below}
\paragraph{Defining Extremal Principal Components}
Given the TPDM $\Sigma$, and its eigenvectors $\{\mathbf{a}_j\}_{j = 1}^D$, we define the \textit{extremal principal components} as $$Z_j = \mathbf{a}_j\T \mathbf{X}, \quad j = 1, \dots, D.$$\index{Dimension reduction in multivariate extremes!extremal principal component} 
Since extremal principal components are linear combinations of the elements of a jointly regularly varying random vector, they are jointly regularly varying with index $\alpha$.
Using the same normalizing function $b(t)$ as was used for $\mathbf{X}$, let $H_\mathbf{Z}$ be the angular measure of $\mathbf{Z} = (Z_1, \ldots, Z_D)\T$.
\cite[Proposition 6]{CT} shows that principal components $Z_i$ and $Z_{j}$, for $i \neq j$, are orthogonal in the sense that $\int_\sphere w_i w_{j} \text{d}H(\mathbf{w}) = 0$.
However, this does not imply that the principal components are asymptotically independent.
%, but does imply \as{"their joint mass is `balanced' in both positive and negative orthants": sentence not so clear to me. Maybe, simply say that  $Z_i,Z_j$ are uncorrelated? I wonder whether wikipedia picture \texttt{https://en.wikipedia.org/wiki/Correlation}  (top-right picture, middle column, last row, the U-shape) is a counter-example to your idea of 'balancing' }.
If $\alpha = 2$, the scale of the individual principal components $\lim_{u \rightarrow \infty} b(u)\,\PP(|Z_j| > u) = \lambda_j$, the $j${th} eigenvalue of $\Sigma$ \cite[Proposition 6]{CT}.

Retaining only the first $d$ eigenvectors, the random vector $\tilde{\mathbf{X}} = \sum_{j = 1}^d Z_j \mathbf{a}_j \in V_{\nu}$ is the optimal $d$-dimensional reconstruction in the sense of \S \ref{sec:probaSetting}, and is regularly varying.
Given observations $\{\mathbf{x}_i\}_{i = 1}^n$, and eigenvectors $\hat{\mathbf{a}}_1, \ldots, \hat{\mathbf{a}}_p$ from the PACE algorithm, and observed principal components $z_{i,j} = \hat{\mathbf{a}}_j\T \mathbf{x}_i$, for $j = 1, \ldots, p$,  $\tilde{\mathbf{x}}_i = \sum_{j = 1}^d z_{i,j} \hat{\mathbf{a}}_j$ can be seen as the best $p$-dimensional reconstruction of the $i$th observation.

%%%%%%%%%%%%%%%%%%%%%%%%%%%%%%%%%%%%%%%%%%%%%%%%%%%%%%%%
\subsection{The $\tau$-Mapping for Modeling on the Positive Orthant} \label{sec:tau}\index{Dimension reduction in multivariate extremes!$\tau$-mapping} 
\vspace{-0.5cm} \strut \hfill \textcolor{gray}{\footnotesize \cite{CT}}
%paragraph on why the positive orthant
%because characteriazation of MEVD's has support on positive orthant, becomes setting for threshold exceedances.
%direction of risk, can be oriented so 
%TPDM and summarizing
%The PCA method for extremes proposed by \cite{CT} is specifically developed for nonnegative regularly varying random vectors. 
\paragraph{Context}
As evident from previous chapters, it is common in multivariate extreme value theory to assume that the random variable has support on the positive orthant $(0, \infty)^D$. The positive orthant may be the natural support of the variables being studied, as is the case for the rainfall data depicted in the right panel of Fig.~\ref{fig:classicalPCA} and to be studied in \S \ref{sec:appRainfall}.
Modeling in the positive orthant can also be motivated by classical extreme value theory, where a common characterization of the multivariate extreme value distributions assumes that the marginals  after standardization have Fr\'echet distributions  with support on $(0, \infty)$.
As an extreme value analysis often begins with an assumption that the distribution is in the domain of attraction of an extreme value distribution, the positive orthant can be seen as a natural setting for modeling threshold exceedances.

\paragraph{A Tail-Preserving Mapping}
One may notice a limitation of Algorithm~\ref{algo:PACE} when applied to non-negative data:  interpreting the eigenvectors $\hat{\mathbf{a}}_j$'s is difficult as they contain negative entries.
More precisely, as the data are uncentered, $\mathbf{a}_1$'s elements have a common sign (we orient so that all entries are positive), but $\mathbf{a}_j$, would contain negative entries for $j > 1$.
A second consequence is that the partially reconstructed data---i.e.,  the data projected on $\hat V= \Span(\hat{\mathbf{a}}_1,\ldots, \hat{\mathbf{a}}_d)$---may also contain negative entries, although the probability of such an event is small if $d$ is close enough to $D$, due to the proximity between $\mathbf{X}$ and its reconstructed version. 

To address these issues, one can adopt the approach proposed by \cite{CT}. The overarching idea is to ensure that partial reconstructions lie in $(0, \infty)^D$, while preserving the tail behavior of the data. Although \cite{CT} introduce a rich class of mappings to achieve this, we focus here on a specific variant known as the \textit{softplus activation function},\index{Neural model!activation function!softplus}
\begin{equation*}
  \tau(y) = \log\{1 + \exp(y)\}.
\end{equation*}
It follows that $\tau(y)$ is \textit{tail equivalent} to $y$, in the sense that $\tau(y) / y \to 1$ as $y \to \infty$, which implies that the softplus transformation has vanishing influence in the tail; the same applies to its inverse $\tau^{-1}(x) = \log\{\exp(x) - 1\}$. This activation function can be used to define `sum' and `scalar' multiplication operations that are essential for constructing partial reconstructions:
\begin{equation}\label{ch11ops}
\mathbf{x}_1 \oplus \mathbf{x}_2 = \boldsymbol{\tau}\left(\boldsymbol{\tau}^{-1}(\mathbf{x}_1) + \boldsymbol{\tau}^{-1}(\mathbf{x}_2)\right), \qquad
a \circ \mathbf{x}_1 = \boldsymbol{\tau}\left(a \, \boldsymbol{\tau}^{-1}(\mathbf{x}_1)\right),
\end{equation}
where $\mathbf{x}_1, \mathbf{x}_2 \in (0, \infty)^D$, $a \in \mathbb{R}$, $\boldsymbol{\tau}(\mathbf{x}) = (\tau_1(x_1), \dots, \tau_D(x_D))\T$, and $\boldsymbol{\tau}^{-1}$ is defined analogously.

\paragraph{Connections between PACE and \cite{CT}}
The extremal PCA method proposed by \cite{CT} is strongly connected to the PACE algorithm above.
The method in \cite{CT} starts with an estimate for the TPDM $\hat \Sigma$ and obtains the same eigenvectors $\{\hat{\mathbf{a}}_j\}_{j = 1}^d$ output from PACE.
However, \cite{CT} view these $\hat{\mathbf{a}}_j$'s as preimages and create the vectors $\hat{\mathbf{e}}_j = \tau(\hat{\mathbf{a}}_j)$ which form the reduced-dimension subspace of $(0,\infty)^D$.
Sample principal components are defined as $z_{i,j} = \hat{\mathbf{a}}_j\T \tau^{-1} (\mathbf{x}_i)$.
Lower dimensional reconstructions $$\tilde{\mathbf{x}}_i = \bigoplus_{j = 1}^d z_{i,j}\circ \mathbf{e}_j$$ which are based on the operations in \eqref{ch11ops}, are guaranteed to be in the positive orthant.\footnote{A measure-theoretic argument, beyond the scope of this book, also shows that under mild conditions one can first use PACE to obtain the optimal $\tilde{\mathbf{X}}$ in $\rset^D$, and subsequently map it onto $(0, \infty)^D$ while preserving its tail structure. This result applies not only to the softplus activation function, but to the entire class of functions introduced in \cite{CT}.}

The workflow proposed by \cite{CT} may thus be described as follows:

%\begin{shortbox}\vspace{-0.3cm}
  \textbf{Extremal PCA: Interpretation and Representation}
  \begin{enumerate}
  \item Run Algorithm~\ref{algo:PACE} (output: $\widehat{\Sigma}, \hat{\mathbf{a}}_1,\ldots \hat{\mathbf{a}}_p$). 
  \item For interpretation purposes: Apply the $\tau$-transform to the eigenvectors, and consider $\mathbf{e}_j = \boldsymbol{\tau}(\mathbf{a}_j)$  for interpretation.
  \item For representation purposes: Propose a low rank representation of $\mathbf{X}$, $\tilde{\mathbf{X}}  = \boldsymbol{\tau}(\sum_{j=1}^p \tilde{Z}_j \mathbf{a}_j)$. %, which corresponds to a transformed linear combination of $\tilde Z$ and $\{e_1, \ldots, e_p\}$.
  \end{enumerate}% \vspace{-0.3cm}

%% file: applicationFinance.tex
\subsection{Financial Data}\label{sec:finData}\index{Dataset!Portfolios from Kenneth R.~French Data Library}
\paragraph{Data Description}
We apply Algorithm~\ref{algo:PACE} to the negative value-averaged daily return data of 30 industry portfolios obtained from the Kenneth R.~French Data Library. We study the period from January 1970--November 2023 inclusive, yielding $n = 13\,599$ observations. This dataset was studied for an earlier period in \cite{CT} and has recently been investigated by  \cite{kiriliouk2022}. Let $\mathbf{x}_i$ be the vector of negative daily returns for day $i$.  

These data appear to be heavy-tailed. A Hill plot\index{Threshold selection!Hill plot} of $\hat \alpha_{(k)}$ (not shown), applied to $\| \mathbf{x}_i \|$, appears to be relatively stable for $k < 600$. With $k = 370$, $\hat \alpha_{(k)} = 3.35$. Hill plots of individual components $|x_{i,j}|$ yield similar values for $\hat \alpha_{(k)}$, leading us to conclude (as did \cite{kiriliouk2022}) that an assumption of a common $\alpha$ for these data is reasonable.

\paragraph{Implementing PACE and Interpretation}
We implement PACE and obtain an eigendecomposition of $\hat \Sigma_k$. The leading five eigenvalues are 0.743, 0.063, 0.033, 0.026, 0.018; recall $\sum_{j = 1}^D \lambda_j = 1$.

\begin{figure}[htb]
\includegraphics[width=14 cm]{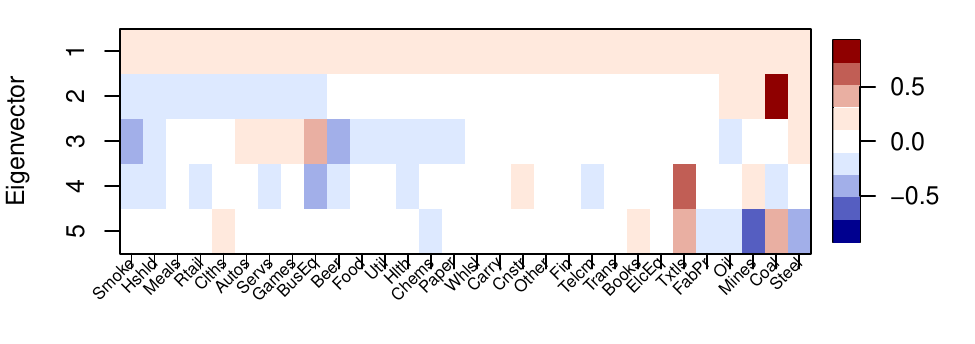}
\caption[Short figure caption]{Leading five eigenvectors $\mathbf{a}_1, \ldots, \mathbf{a}_5$ of $\hat \Sigma_k$ for the 30 industry portfolios data.} \label{fig:eigenvectorsFinance}
\end{figure}

\begin{figure}[htb]\centering 
\includegraphics[width=11 cm]{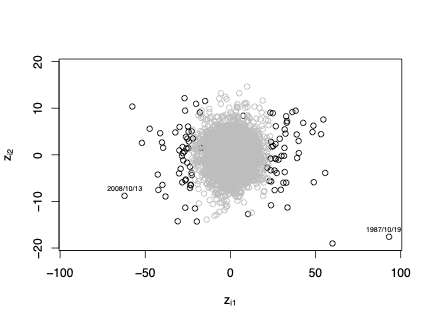}
\caption[Short figure caption]{Biplot of the first two sample principal components. The 100 days with the largest magnitudes are plotted in black.} \label{fig:biplot}
\end{figure}

The sample eigenvectors $\hat{\mathbf{a}}_1, \ldots, \hat{\mathbf{a}}_5$ are pictured in Fig.~\ref{fig:eigenvectorsFinance}.
We define the sample principal components $z_{i,j} = \hat{\mathbf{a}}_j\T \mathbf{x}_i$.
As the leading eigenvector $\hat{\mathbf{a}}_1$ is relatively constant, an interpretation is that the leading principal component $z_{i,1}$ accounts for large movements (positive or negative) in the return across all industries.
After accounting for overall movement across sectors, the second principal component $z_{i,2}$ distinguishes movement between `heavy' industries of steel, mines, coal, and oil, and `lighter' industries including games, household, services, business equipment, and retail.
This interpretation of the eigenvectors is perhaps best seen by examining particularly noteworthy days.
Fig.~\ref{fig:biplot} shows plots of the first two sample principal components, the days with the largest magnitudes $\| \mathbf{x}_i \|$ are plotted in black, and two days are denoted by their date.
The point labeled ``1987/10/19" refers to a day known as ``Black Monday", a day when US markets experienced an unexpected and dramatic drop, which is why the leading sample principal component for these negative returns takes its largest value.
While the overall market experienced huge losses, the negative value of the second principal component mitigates these losses for the aforementioned `heavy' industries:  mines and coal had the smallest losses among these 30 industries.
The point labeled ``2008/10/13" corresponds to a day when the overall market experienced large gains in reaction to worldwide governmental responses to the financial crisis at that time, and the industries with the largest returns from that day were coal, steel, oil, and mines respectively. Fig.~\ref{fig:biplot} clearly shows that $Z_1$ and $Z_2$ are not asymptotically independent, but that the directions of the largest values appear to be uncorrelated between the positive and negative quadrants.

% \cite{DS} illustrate the optimality of the eigendecomposition via a simulation study.  
% Here, as we do not have a true lower-dimensional subspace, we are unable to demonstrate optimality of the lower dimensional subspace given by the eigenbasis of dimension $d$.
% Instead, we compare the risk of the linear subspace given by the first five eigenvectors to an `informed' 5-dimensional subspace obtained by partitioning the 30 industries into 5 categories denoted in a 5-industry portfolio also found in the Kenneth R. French Data Library.   
% Table~\ref{tab:portfolio} gives the partition, which was done by an informal mapping of the 4-digit SIC codes appearing in the details given about each portfolio.
% The basis vectors for this subspace are indicator vectors, normalized to have unit length.
% %\DC{We may want to put this table should be put at the very end of the entire chapter, or stuck somewhere online.  It's at the end of this section for now since it's attached to the "applicationFinance.tex" file.}
% We project the 100 angular components $\theta_i$ corresponding to the observations with largest norm onto the eigenspace and the 5-industry subspace.
% The mean squared error of the eigenspace projection is 0.07, whereas the 5-industry subspace is 0.15.

\paragraph{Further Insights}
PACE provides insights about combinations of portfolios associated with the largest market movements.  
More precisely, for a linear combination $\mathbf{b}\in\mathbb{R}^D$ with unit norm, the dot product
$\mathbf{b}\T \mathbf{X}$ is the movement associated with a portfolio allocated proportionally as $\mathbf{b}$.
and $\mathbf{b}\T \Theta = \mathbf{b}\T \mathbf{X} / \|\mathbf{X}\|$ may be interpreted as the movement of this portfolio relative to overall market movement. 
In case of a shock (represented by $\|\mathbf{X}\|>u$ for large $u$), consider the conditional expectation of the (squared) relative movement, 
$$\EE\{\,(\mathbf{b}\T \Theta)^2\;|\; \|\mathbf{X}\|>u\,\} = \mathbf{b}\T \Sigma_u \mathbf{b}  = 
\sum_{j=1}^D \lambda_{j,u} (\mathbf{a}_{j,u}\T \mathbf{b})^2, $$ 
where $\{(\lambda_{j,u},\mathbf{a}_{j,u})\}_{j = 1}^d$ is the eigendecomposition of $\Sigma_u$. 
The above quantity is largest if $\mathbf{b} = \mathbf{a}_{d,u}$ and smallest if $\mathbf{b} = \mathbf{a}_{d,u}$. %in the direction of the first (\emph{resp.} last) eigenvector. Also, 
%In the limit $t\to\infty$, 
Also, under the assumptions of  Proposition~\ref{prop:concentrationMu},  
and in view of the discussion above the  statement,  %of Proposition~\ref{prop:concentrationMu}, and 
%in the linear span $V_p$ of the first $p$ principal axes  are more likely to be associated with large losses, compared with the ones in the  orthogonal complement  $V_p\T$, wich may be viewed as 'safer'. In particular in view of Proposition 1 
%and in view of the discussion above the statement, 
when $d=d^*$, and $V_{d^*} = V_\nu$,  the probability that a large movement occurs for portfolio combinations $\mathbf{b}$  in $V_{\mu}^\perp$ is negligible compared with the probability of a large loss or profit  for a portfolio in $V_\nu$. 
%The orthogonal complement $V_\nu^\perp$ then contains the "safest" portfolios that are (much) less likely to encounter (very) large losses than portfolios $f\in V_\nu$.}
\paragraph{Focusing on Extreme Losses}
The PCA performed here characterizes market movement in both gain and loss directions, but there could be reason to focus solely on losses.
There is reason to believe that extremal dependence between the market sectors differs for gains and losses; for example, the oil and coal sectors are more strongly tail dependent for losses than for gains.
\cite{CT} focused only on financial losses by converting gains to zeros so that the negative returns would be restricted to the positive orthant, and then used their transformed-linear eigendecomposition.
%\cite{CT} also performed a marginal transformation in order to assume $\alpha = 2$.
When $\mathbf{X}$ is a random vector of negative log-return, the dot
product $\mathbf{b}\T \mathbf{X}$ becomes the loss associated with the portfolio allocated according to $f$, and the orthogonal complement $V_\nu^\perp$ then contains the `safest' portfolios that are much less likely to encounter very large losses than portfolios $\mathbf{b}\in V_\nu$.

%% file: applicationRainfall.tex
\subsection{Precipitation Data}
\label{sec:precip}

\paragraph{Data Description}
In this section, we use extremal PCA to explore extreme precipitation for the Pacific Coast region of Washington and Oregon, USA.
In cooler months, a weather phenomenon known as atmospheric rivers (ARs) refers to narrow bands of concentrated water vapor being transported from the tropics to the Pacific coast of North America.
Intense atmospheric river events can cause destructive flooding, and ARs also play a critical role in water supply for the western US.

\begin{figure}[htb]
\centerline{
\includegraphics[height = 8 cm]{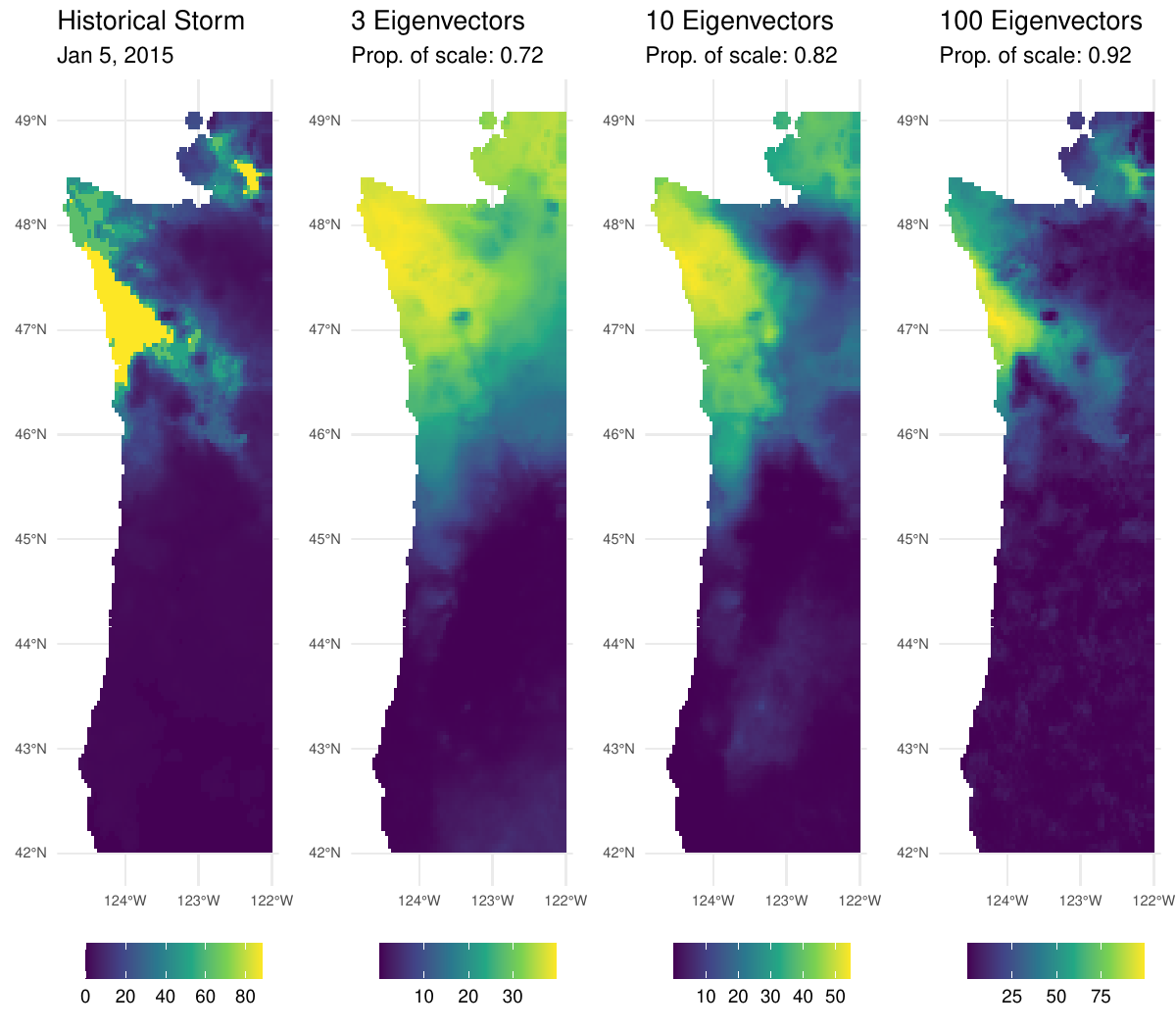}
}
\caption{Left:  Precipitation amounts for January 5, 2015.  The remaining three panels from left to right give partial reconstructions from 3, 10, and 100 eigenvectors respectively.} \label{fig:stormDecomp}
\end{figure}

We apply our extremal PCA method to a precipitation data product from 8510 locations in the Pacific Coast region of Washington and Oregon USA.
Daily data are obtained from PRISM\footnote{\url{https://prism.oregonstate.edu}} in the time period spanning January 1, 1981-March 31, 2023.
To reduce seasonality and to focus on the cooler months, we only retain data for the months between October and March, resulting in a data set with 7744 days.
While our exploration is motivated by ARs, we study all daily precipitation data and do not limit the analysis to known AR events.
We will let $X_i$ denote the random vector of precipitation from these locations on day $i$, and $x_i$ will denote the corresponding observation.
The left panel of Fig.~\ref{fig:stormDecomp} shows precipitation from January 5, 2015, a day corresponding to a known atmospheric river event which caused flooding in coastal communities in Washington state.

\paragraph{Rationale for the Analysis}
From a statistical point of view, there are three interesting aspects about this data set.
First, the dimension is very large.  
Existing multivariate and spatial extremes models are generally very difficult to apply to data sets this large, but PCA can readily be used to explore data sets of this size.
Second, the tail is not very heavy. 
Preliminary analysis fitting a generalized Pareto model to threshold exceedances above the location-wise 0.975 quantile with location-wise scale parameters and a common tail index yields an estimate of $\hat \alpha = 1/\hat \xi = 32$.
That the tail of precipitation data is not estimated to be heavier is somewhat surprising, and could arise from the fact that these are not weather station observations but instead result from a method to interpolate weather station and other data to a high resolution grid.
Third, as indicated in Fig.~\ref{fig:stormDecomp}, many of the extreme events will be localized in comparison to the spatial extent of the study region.
This is different behavior than that of the financial data studied above, where days with large movements in the returns are seen in most, if not all, of the 30 industries.

\paragraph{On TPDM Estimation}
As the tails of these data are not exceptionally heavy and since extreme events are often localized, we alter our procedure for estimating the TPDM.
As is often done in extreme analyses, we use a rank transform to transform the data at each location to have a common marginal distribution $\text{P}(X_i \leq x) = \exp(-x^{-2})$.
This transformation results in the data having a much heavier tail, highlighting extreme events.
The norm of the original precipitation data for January 5, 2015 only corresponds to the 0.57 quantile, but after transformation, this day's norm is the sixth largest in the data set.
Not only does this transformation result in transformed data with a much heavier tail than the original, it yields a common scale for each location.
Having a common scale is analogous to performing standard PCA on a correlation matrix, and choosing whether to perform PCA on a correlation or covariance matrix is a practitioner decision based on understanding of the data and the questions of interest.
Below, $\mathbf{X}_i$ denotes day $i$'s precipitation vector after transformation, and Fig.~\ref{fig:stormDecomp} depicts transformed precipitation as well.

\paragraph{A Problem-Driven Choice of the Norm}
Because extreme events have local extent, we also choose to use \textit{pairwise norms} $\| (x_{i,j}, x_{i,j'}) \|$ rather than \textit{vector norms} $\| \mathbf{x}_i \|$, to select threshold exceedances to estimate $\sigma_{j,j'}$.
In theory, the choice does not matter as $H$ captures extremal dependence as $X$ becomes large, no matter how `large' is defined.
To understand the motivation for using pairwise norms, consider a day $i$ where a storm causes $\| \mathbf{x}_i \|$ to be large, but two adjacent locations $j$ and $j'$ lie outside the storm's spatial extent.
The estimate arising from (\ref{eq:tpdmk}) would include day $i$, but day $i$'s contribution to $\sigma_{j,j'}$ would be near zero as both $\theta_{i,j}$ and $\theta_{i,j'}$ would be near zero.
The resulting estimator could then be negatively biased.
\cite{mhatre2023} shows how the TPDM of $X$ is related to the angular measure of its bivariate marginals, providing us the estimate 
$$\hat \sigma_{j,j'} = c \frac{1}{k} \sum_{i = 1}^n \frac{x_{i,j}}{\| (x_{i,j}, x_{i,j'}) \|}  \frac{x_{i,j'}}{\| (x_{i,j}, x_{i,j'}) \| } I(\| (x_{i,j}, x_{i,j'})  \| \geq \| (x_{i,j}, x_{i,j'}) \|^{(k)}).$$
Because we work with standardized data, $c = 2$.
$\hat \Sigma = [\hat \sigma_{j,j'}]_{j,j' = 1, \ldots, D}$ is not guaranteed to be positive definite, so our final TPDM estimate $\tilde \Sigma$ is obtained by applying a method from \cite{higham2002} to obtain the nearest positive definite matrix to $\hat \Sigma$.

\paragraph{Analysis}
Extremal PCA begins by taking the eigendecomposition of $\tilde \Sigma$.
As $\alpha = 2$, the diagonals of the TPDM have the scale interpretation given in \S \ref{sec:tau}.
The leading eigenvalue accounts for 55\% of the total scale, and the first three eigenvalues account for 72\% of the total scale.
However, because of the localized nature of extreme events, it takes a large number of eigenvalues to capture a large proportion of the scale; the first 20 eigenvalues account for 86\%, the first 100 for 92\% and the first 1000 for 96\%.
Perhaps contrary to intuition, this can give even more motivation for performing a PCA analysis.
The directions of the leading eigenvectors provide useful information about the large-scale behavior or trends of extreme precipitation, which can be hard to investigate directly from the data due to the extremes' localized behavior.

\begin{figure}[htb]
\centerline{
\includegraphics[height = 8 cm]{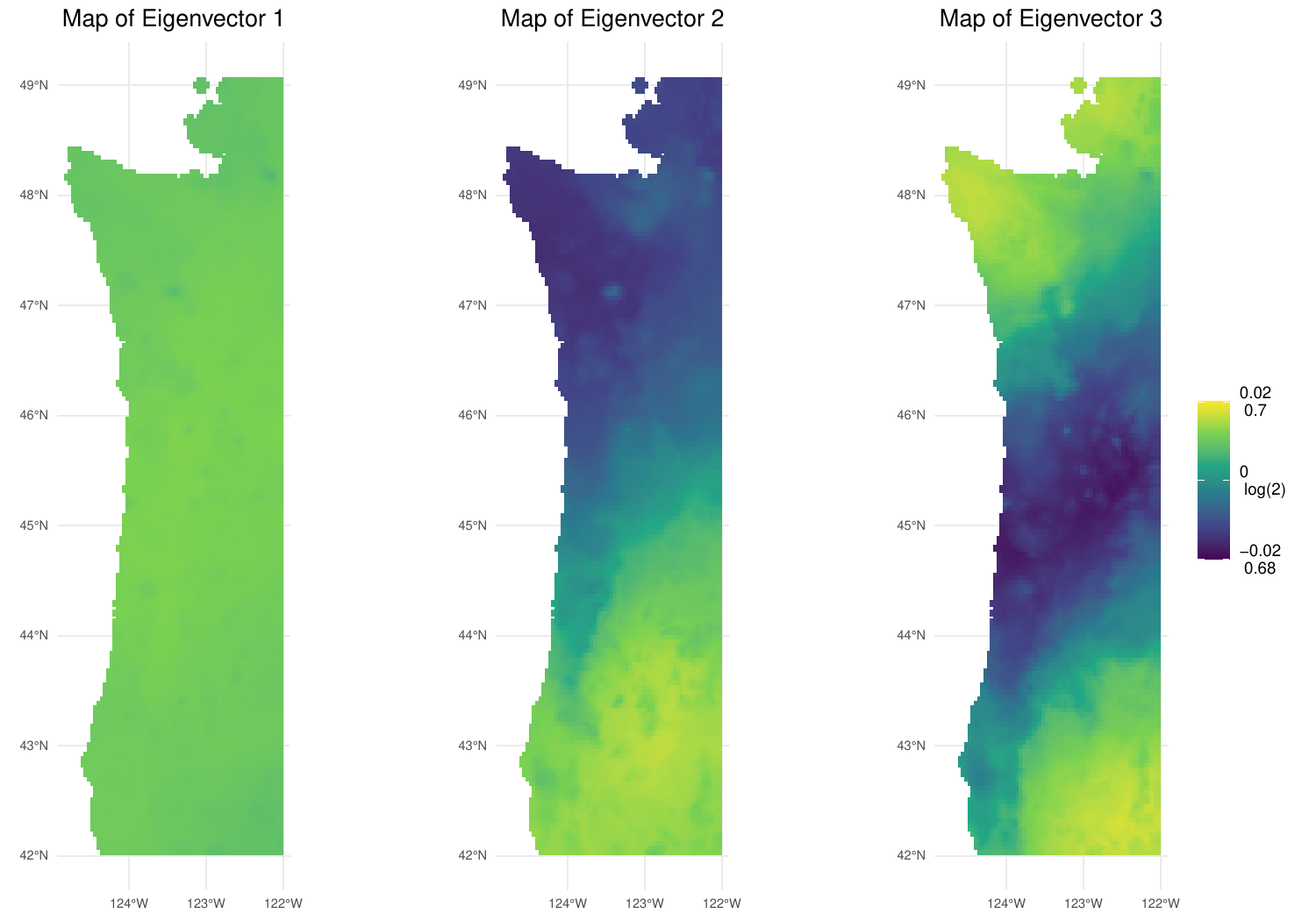}
}
\caption{Leading three eigenvectors from the rainfall data.}\label{fig:eigenvectorsRainfall}
\end{figure}

As precipitation is nonnegative, $\mathbf{X}_i$ take on values in $[0, \infty)^D$ and we will choose to use the basis of \cite{CT} $\hat{\mathbf{e}}_j = \tau(\hat{\mathbf{a}}_j)$, where $\hat{\mathbf{a}}_j$ is the $j$th eigenvector of $\tilde \Sigma$. Fig.~\ref{fig:eigenvectorsRainfall} shows the first three eigenvectors.  
The color scale is given in terms of both $\hat{\mathbf{e}}_j$ and $\hat{\mathbf{a}}_j$ for interpretability.
The first eigenvector is relatively constant and $\hat{\mathbf{e}}_{1,k} > 0_{\mathbb{R}^D} = \log(2)$ for all $k$, so it can be interpreted as providing information about the magnitude precipitation vector for the study region.
The second eigenvector takes on positive values (i.e., $\hat{\mathbf{e}}_{2,k} > 0_{\mathbb{R}^D}$) in the southern portion of the study region, and negative values in the north; thus eigenvector two allocates precipitation to the north or the south after eigenvector one accounts for overall magnitude..
The third eigenvector has negative values in the center of the study region and positive values in both the north and south, thus this eigenvector further apportions the precipitation regionally. 

We define the principal component scores as $z_{i,j} = \boldsymbol{\tau}^{-1}(\hat{\mathbf{e}}_j)\T \mathbf{x}_i = \mathbf{a}_j\T \mathbf{x}_i$.
Any observation $x_i$ can be partially reconstructed in $\mathbb{R}^d_+$ as a transformed linear combination of the leading $p$ eigenvectors: $\tilde{\mathbf{x}}_i =  \boldsymbol{\tau}(\sum_{j = 1}^d z_{i,j} \mathbf{a}_j)$. %y_{i,1} \circ e_1 \oplus \ldots, \oplus y_{ip} \circ e_p$.
Three panels of Fig.~\ref{fig:stormDecomp} give partial reconstructions for $d = 3, 10, 100$.
The $d = 3$ reconstruction illustrates that the leading few principal components capture (only) the large scale behavior as this partial reconstruction indicates heavy precipitation broadly in the northern area of the study region.
As also indicated by the proportion of scale given by the eigenvalues, we see that some of the fine detail of an individual storm is still not completely accounted for by the $d = 100$ reconstruction.

\begin{figure}[htb]
\centerline{
\includegraphics[width = 12 cm]{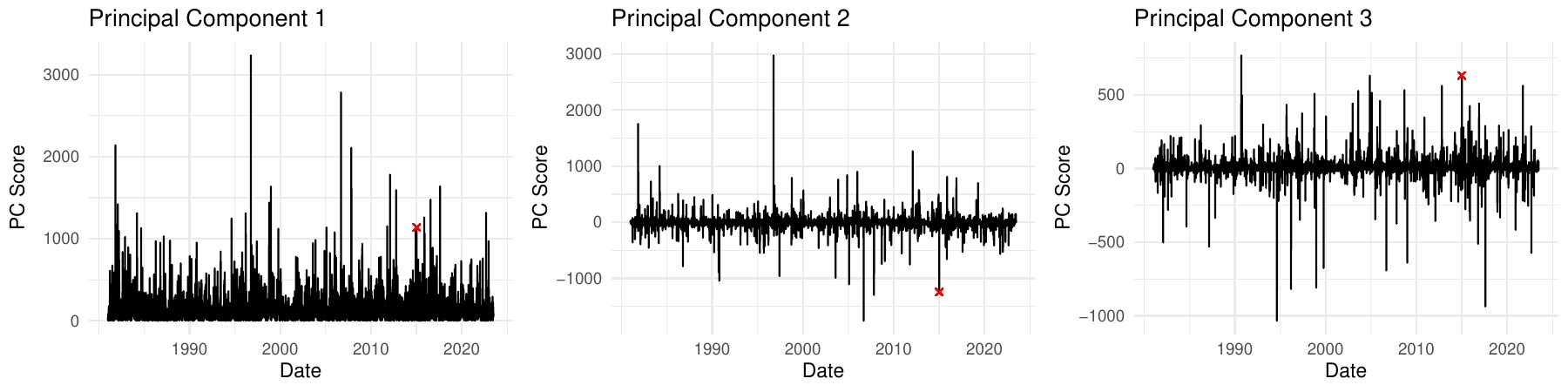}
}
\caption{Time series plots of the scores for principal components 1, 2, and 3.  The storm of January 5, 2015 is denoted with a red `x'.}\label{fig:pcts}
\end{figure}

\paragraph{Takeaways from the Analysis}
Fig.~\ref{fig:pcts} shows the time series of the scores for principal components 1, 2, and 3, and the date corresponding to January 5, 2015 is denoted by a red `x' in each plot.  
Questions about large scale behavior can be investigated via these time series.
For instance, there is widespread interest in understanding if extreme precipitation events are becoming more intense due to climate change.
Such questions can be difficult to investigate directly from data such as these; for example,  an estimated trend in marginal parameters would likely appear spatially noisy due to the local nature of extreme events.
As the score $z_{i,1}$ can be interpreted as a measure of the overall magnitude for precipitation for day $i$, we can investigate for a trend in these scores.
Following \cite{jiang2020}, we perform quantile regression at the 0.975 quantile of $z_{i,1}$, but find the slope is insignificantly different from zero when a hypothesis test is performed with a 0.05 error rate.
This finding is in contrast to \cite{jiang2020} who found a significant positive trend in the leading principal component scores for extreme precipitation.
The study differs in important ways from \cite{jiang2020}, who studied data from weather stations across the continental United States, and focused on different months corresponding to the Atlantic hurricane season.
\cite{jiang2020} also found that principal component scores whose eigenvectors were connected to Atlantic coast precipitation had significant relationships with the El Ni\~no Southern Oscillation index.
%\textcolor{red}{For this Pacific Coast data, we have not found any significant relationships between these principal component scores and climatological drivers such as the WP}.

\begin{figure}[htb]
\centerline{
\includegraphics[width = 4 cm]{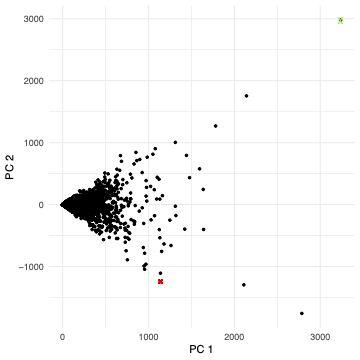} \hspace{2 cm}
\includegraphics[width = 4 cm]{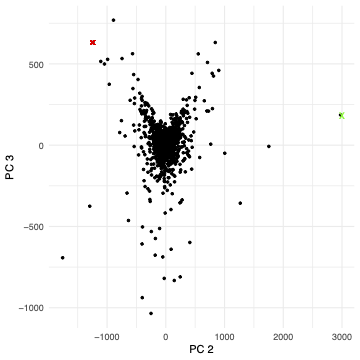}
}
\caption{Biplots of principal component (PC) scores. The storms of January 5, 2015 and November 19, 1996 are denoted with a red and green `x' respectively.} \label{fig:biplots}
\end{figure}

%Another exploratory tool arising from a principal component analysis are biplots of principal component scores.
%Studying the financial industry portfolio data, \cite{CT} found an interesting direction of extreme behavior in the biplot of the scores corresponding to the second and third principal components.
%The day with the largest magnitude in this direction was April 2, 2002, which is known as `Marlboro Friday' in the financial literature.
Fig.~\ref{fig:biplots} shows biplots of the scores the first three principal components.
The right panel shows three primary modes of directions for the scores with large magnitude: a downward direction correponding to large negative values of $z_{i,3}$ and relatively small values of $z_{i,2}$, a right-upward direction in which both $z_{i,2}$ and $z_{i,3}$ are large and positive, and a left-upward direction in which $z_{i,2}$ is large and negative and $z_{i,3}$ is large and positive.
Combining these directions with the eigenvectors given in Fig.~\ref{fig:eigenvectorsRainfall} yields the interpretation that the downward direction corresponds with large precipitation events in the center of the study region, right-upward corresponds with large precipitation in the south, and left-upward corresponds with large precipitation in the north.
One exploratory use of principal component biplots is to identify outliers.
Of note in the right panel of Fig.~\ref{fig:biplots} are a small number of points with large magnitude which do not seem to correspond with the three primary modes of direction.
Denoted in this panel by a green 'x' is the day of November 19, 1996, which appears to be an outlier in the biplot since  its direction does not correpond to any of the three modes.
Upon investigation, what appears to be unique about this event is its spatial extent; a map of this day's precipitation over our study region shows very high precipitation extending from the study region's southern border to about 45$^\circ$ N latitude.
Beyond our study region, there are reports of flooding not only in Oregon, but also in areas of the adjacent states of California and Nevada.
An account of this storm\footnote{\url{https://pubs.usgs.gov/fs/2004/3134/}} describes the precipitation arising from a ``broad upper-air weather system," which appears to be rather unique in this data record.

%
%
%
%
%
%%%%%%%%%%%%%%%%%%%%%%%%%%%%%%%%%%%%%%%%%%%%%%%%%%%%%%%%%%%%%%%%%%%%%
%%%%%%%%%%%%%%%%%%%%%%%%%%%%%%%%%%%%%%%%%%%%%%%%%%%%%%%%%%%%%%%%%%%%%
%%%%%%%%%%%%%%%%%%%%%%%%%%%%%%%%%%%%%%%%%%%%%%%%%%%%%%%%%%%%%%%%%%%%%
%
%
%
%\mainmatter
%
%\bibliographystyle{plain}
%\bibliography{bibtexPCA}
%
%\printindex
%\cleardoublepage
%
%\end{document}

%% file: perspectives.tex
There are several open directions for further exploration regarding PCA for extremes.
From a theoretical perspective, several questions remain unanswered. First,  the statistical properties of the method based on the $\tau$-transform proposed in~\cite{CT} have not been investigated. Regarding the PACE  algorithm~\ref{algo:PACE} itself (without the $\tau$-transform), the guarantees derived by~\cite{DS} are likely to be suboptimal in terms of excess of reconstruction risk. In fact~\cite{blanchard2007statistical} were able to obtain fast rates (of order $1/n$), which suggests that under appropriate assumptions, in an EVA context,  the excess risk of the empirical risk minimizer should be of order $O(1/k)$. Finally, no  asymptotic analysis has been carried out yet, which would provide insights upon the tightness of the bounds obtained in~\cite{DS}. 

In a different direction,  the increasing availability of functional data in ever larger databases suggests extending the PACE framework to functional spaces. This is the main motivation behind the recent  paper~\cite{clemenccon2023regular} which develops a theoretical framework for  functional PCA of extreme $L^2$ valued random functions,  with a focus on theoretical guarantees paralleling the ones obtained in \cite{DS} with different techniques of proofs accommodating for an infinite dimension,  opening avenues for novel applications involving  irregularly sampled or high frequency data.

In this chapter (as in \cite{DS,clemenccon2023regular}, and \cite{CT}),  we solely considered uncentered second moment matrices $\EE(\boldsymbol{\Theta}\boldsymbol{\Theta}\T \; | \;  \|\mathbf{X}\|>u)$ and their limit as $u\to\infty$. A natural extension would be to allow for a centering step and consider the covariance matrix of $\Theta$ conditioned upon a radial exceedance, say $\EE\{(\boldsymbol{\Theta}- \mathbf{m}_u)(\Theta - \mathbf{m}_u)\T \;|\; \|\mathbf{X}\|>u\}$, where $\mathbf{m}_u = \EE(\boldsymbol{\Theta} \;| \; \|\mathbf{X}\|>u)$.  In view of the earlier discussion regarding uncentered PCA, one  may anticipate that this should remove the first principal component typically observed with (extreme or not extreme) non-centered PCA, see \cite{cadima2009relationships}. From a mathematical statistics perspective, as $\mathbf{m}_u$ is unknown, including centering term  $\hat{\mathbf{m}}_u$ would  introduce an additional layer of technicality as the random variables $(\Theta_i - \hat{\mathbf{m}}_u)_{i\le n}$ would not independent. Such issues are typically handled by means of U-statistics and their Hoeffding decompositions, which is precisely the path followed in \cite{blanchard2007statistical} (\S 3.5) for traditional and kernel centered PCA. To date, obtaining guarantees for a centered version of extremal PCA remains an open problem. For further discussion, see Remark 4.5 in \cite{clemenccon2023regular}.

In terms of application, here we have presented PCA as a tool for exploring high dimensional extremes.
However, PCA can also be viewed as a way for constructing a lower-dimensional model for extremes.
Factor analysis is a non-extreme method for representing multivariate data in terms of a smaller number of latent variables which are related to the data via factors, which usually begin as the PCA eigenvectors, and which may be then rotated for interpretabilty.
Recent work by \cite{rohrbeck2023} used the transformed eigenvectors of \cite{CT} to build a lower dimensional model for flood events measured gauging stations in Northern England and Southern Scotland.
Rather than directly modeling the data from the 45 stations, \cite{rohrbeck2023} constructed a regular varying model for the leading six principal components.
The principal components $Z_j$ are `uncorrelated' in the sense that off-diagonal elements of their TPDM are zero \cite{CT}, and this is seen in Fig.~\ref{fig:biplots} as integration with respect to the angular measure will `balance' in positive and negative directions.
However, as Fig.~\ref{fig:biplots} also clearly shows, the principal components are not asymptotically independent, and thus cannot be trivially modeled.
\cite{rohrbeck2023} semiparametrically modeled the angular measure of the leading principal components, and also modeled a relationship between these principal components and the remaining part of $\mathbf{X}$ not accounted for by these principal components (which is called the error or specific factors in factor analysis).
\cite{rohrbeck2023} then used the lower dimensional model to produce a large number of synthetic storms and verified that the extremal dependence in these synthetic storms closely matched that found in the data.
Importantly, the study region in \cite{rohrbeck2023} was not large and much of the overall scale in the original data was accounted for by these leading principal components.
Lower-dimensional modeling via PCA would be more challenging for data such as the rainfall data above since reconstructions require so many eigenvectors to produce realistic storm characteristics.

As a final remark, applications of the PCA methodology presented in this chapter have no reason to be limited to financial or environmental sciences settings.  % may  in principle open the road to to
In particular novel applications to  anomaly detection\index{AI and Machine Learning!anomaly detection} among extremes  could likely be devised based on the PACE algorithm, leveraging ideas from   on the one hand, \cite{goix2016sparse,goix2017sparse} who propose generic approaches for such purposes based on the construction of a scoring function, see also~\cite{clemenccon2023concentration} ; and on the other hand, classical approaches of anomaly detection  based on PCA coming from the  Computer Science literature (\cite{camacho2016pca,NIPS2006_2227d753}).

%%% Local Variables:
%%% mode: latex
%%% TeX-master: "template"
%%% End: